\documentclass[aps,prd,reprint,nofootinbib,longbibliography,showkeys,dvipsnames, twoclumn]{revtex4-1}
\usepackage{ORCIDinREVTeX}

\usepackage{amsmath, amsfonts, amssymb, autobreak, cancel, mathtools, amscd, amstext, graphics, float, braket, graphicx, multirow, contour,ragged2e,soul,slashed,comment,multirow}
\usepackage{xcolor}

\usepackage{cellspace} 
\usepackage{url}
\usepackage{adjustbox}
\usepackage[colorlinks=true,urlcolor=blue,linkcolor=blue,citecolor=blue,anchorcolor=blue,filecolor=blue,menucolor=blue,linktocpage=true,pdfproducer=medialab,pdfa=true]{hyperref}
\usepackage[nameinlink]{cleveref}

\usepackage{etoolbox}

\pretocmd{\cref}{\textbf{}}{}{}
\pretocmd{\Cref}{\textbf{}}{}{}

\crefname{equation}{Eq.}{Eqs.}
\Crefname{equation}{Eq.}{Eqs.}
\Crefname{footnote}{footnote}{footnotes}
\newcommand{\appref}[1]{\hyperref[#1]{Appendix~\ref*{#1}}}
\newcommand{\appsref}[2]{\hyperref[#1]{Appendices~\ref*{#1}~and~\ref*{#2}}}

\usepackage[compat=1.1.0]{tikz-feynman}

\newcommand{\cMH}[1]{\textcolor{purple}{\textbf{[#1]}}}

\usepackage{slashed}

\begin{document}
\author{Juan~P.~Garc\'es}
\orcid{0000-0002-6933-8750}
\email{juan.garces@mpi-hd.mpg.de}
\affiliation{Max-Planck-Institut für Kernphysik, Saupfercheckweg 1, 69117 Heidelberg, Germany}
\author{Florian~Goertz}
\orcid{0000-0001-8880-0157}
\email{florian.goertz@mpi-hd.mpg.de}
\affiliation{Max-Planck-Institut für Kernphysik, Saupfercheckweg 1, 69117 Heidelberg, Germany}
\author{Maya~Hager}
\orcid{0009-0003-7830-5007}
\email{maya.hager@mpi-hd.mpg.de}
\affiliation{Max-Planck-Institut für Kernphysik, Saupfercheckweg 1, 69117 Heidelberg, Germany}
\title{Dirac leptogenesis with charged leptons}
\begin{abstract}
    Popular frameworks that explain the baryon asymmetry of the Universe (BAU) are electroweak baryogenesis, requiring a first-order phase transition, and Majorana leptogenesis, which relies on lepton number violation in the neutrino sector. Dirac leptogenesis offers an alternative mechanism that avoids both requirements, instead allowing the BAU to follow from previously generated chiral asymmetries that remain out of equilibrium in the early Universe. Conventional realizations of this scenario typically invoke ultra-light Dirac neutrinos. In this work, we generalize the mechanism to include Standard Model charged leptons by assuming that their mass-generating interactions have a UV origin and arise from higher-dimensional operators. For concreteness, the analysis is performed within a Froggatt–Nielsen setup, which allows us to naturally address fermionic mass hierarchies, too, now linked to the BAU, and we show that the proposed mechanism is viable by supercooling the flavon phase transition. The framework predicts a flavon with a $(1.7 - 50)\,\mathrm{GeV}$ mass, bounded by the non-observation of lepton flavor violation and the requirement for the Universe to not reheat above sphaleron temperatures, which could be indirectly probed by future lepton flavor violation experiments.
\end{abstract}

\maketitle

%
\section{Introduction}
The observed baryon asymmetry of the Universe 
(BAU)~\cite{Planck:2015fie,Cooke:2017cwo,Planck:2018vyg,Yeh:2022heq} cannot be explained within the current version of the Standard Model (SM) of particle physics, which motivates the search for new particles and interactions. Since Sakharov's foundational work~\cite{Sakharov:1967dj}, numerous mechanisms have been proposed to explain this matter-antimatter asymmetry. Among the most extensively studied are electroweak baryogenesis~\cite{Kuzmin:1985mm,Shaposhnikov:1986jp,Farrar:1993hn} and Majorana leptogenesis~\cite{Fukugita:1986hr,Luty:1992un,Covi:1996wh}. The former requires a sufficiently strong first-order electroweak phase transition (FOPT), while the latter relies on lepton-number violation (LNV) in the neutrino sector, neither of which is provided by the SM.

Dirac leptogenesis~\cite{Dick:1999je,Akhmedov:1998qx} offers an alternative based on Dirac neutrinos that requires neither a FOPT nor LNV in the neutrino sector. In its standard realization, the SM is supplemented with right-handed neutrinos, $\nu_R$, with tiny Yukawa couplings that account for the observed light neutrino masses~\cite{Super-Kamiokande:1998kpq,SNO:2002tuh}. As shown in Ref.~\cite{Dick:1999je}, these feeble interactions prevent left-right equilibration, allowing a primordial chiral asymmetry in the neutrino sector to remain \textit{sequestered} until after electroweak (EW) sphalerons freeze out. While active, sphalerons partially convert the left-handed lepton asymmetry into baryon number, leaving a nonzero BAU that survives to the present day.

The simplicity and minimalism of Dirac leptogenesis have motivated numerous studies connecting the origin of the BAU via this mechanism to other open questions in particle physics and cosmology. These include Dirac neutrino mass generation through minimal ultraviolet (UV) completions~\cite{Murayama:2002je,Gu:2006dc,Cerdeno:2006ha}, Dirac seesaws~\cite{Gu:2006dc}, and radiative mechanisms~\cite{Borah:2016zbd}, as well as connections to the flavor puzzle~\cite{Chen:2011sb,Singh:2024imk}, dark matter~\cite{Borah:2016zbd,Dutta:2022knf,Berbig:2023uzs,Arcadi:2026szf}, inflation and reheating~\cite{Abel:2006hr,Berbig:2022pye,Barrie:2024yhj,Ahmed:2025vzl}, and the strong CP problem~\cite{Gu:2016hxh,Berbig:2022pye,Babu:2024glr}. Since the original proposal~\cite{Dick:1999je} leaves the source of the primordial chiral asymmetry unspecified, several production mechanisms have also been explored~\cite{Abel:2006hr,Berbig:2023uzs,Blazek:2024efd,Ahmed:2025vzl}.

Despite their diversity, these realizations predominantly retain the central ingredient of the original mechanism, namely small fundamental or effective Dirac-neutrino Yukawa couplings that keep the right-handed neutrino asymmetry sequestered until after sphaleron freeze-out. Even when these couplings originate from a UV flavor symmetry or seesaw-like constructions, the interactions relevant during this epoch are typically already well described by the low-energy effective theory with the feebly coupled $\nu_R$ sector serving as the asymmetry reservoir.

A different possibility was explored in Ref.~\cite{Chen:2019wnk}, where the chiral reservoir is provided instead by the right-handed electron. In this scenario, the generated electron asymmetry remains successfully sequestered until sphaleron freeze-out because a non-standard cosmological evolution increases the Hubble rate, thus delaying the equilibration of electron Yukawa interactions. This demonstrates that the essential ingredient of Dirac leptogenesis is not necessarily the presence of a right-handed neutrino, but rather the existence of a chiral charge whose equilibration is sufficiently delayed.

In this work, we show that the chiral asymmetry sequestering mechanism of Dirac leptogenesis can also be extended beyond the right-handed neutrino reservoir, without a modified cosmological history above sphaleron freeze-out as a necessary condition. We demonstrate that this is realizable when giving a UV origin to the Yukawa couplings by considering fermion mass-generating interactions that remain higher-dimensional before sphaleron freeze-out, i.e. without yet reducing to effective dimension-four Yukawa couplings. In this case, the corresponding chirality-changing rates can be parametrically different from those naively inferred from the observed low-energy fermion masses. The same exponential suppression that explains fermion hierarchies also leads to a suppression of the equilibration rate. As a consequence, the right-handed components of other SM leptons, which are already present in the theory, can act as asymmetry reservoirs. We demonstrate this mechanism within a Froggatt-Nielsen (FN) framework~\cite{Froggatt:1978nt} in which the FN symmetry remains unbroken until after sphaleron freeze-out.

The paper is structured as follows. In~\Cref{sec:DLinanutshell} we review the working principle of Dirac leptogenesis, and identify the standard requirement to successfully sequester the lepton asymmetry until sphaleron freeze-out. In~\Cref{sec:DLinFN}, we show how higher-order operators that generate fermion masses in an FN setup allow extending the leptogenesis framework beyond neutrinos, as long as the FN symmetry remains unbroken until after sphaleron freeze-out. We review the relevant constraints from flavor-violation observables in~\Cref{sec:flavObs} and from the supercooling of the FN phase transition in~\Cref{sec:PT}. Finally, we conclude in~\Cref{sec:conclusion}.
%
\section{Dirac leptogenesis in a nutshell}\label{sec:DLinanutshell}
The sequestering mechanism underlying Dirac leptogenesis~\cite{Dick:1999je} is based on the fact that EW sphalerons act only on fields charged under $SU(2)_L$. While active, sphalerons partially convert a left-handed neutrino asymmetry ($\Delta L_L$) into a baryon asymmetry, with a compensating lepton asymmetry ($\Delta L_R$) remaining sequestered in right-handed neutrinos. This process can take place with global lepton number remaining conserved ($\Delta L = \Delta L_L + \Delta L_R = 0$) during the generation of the neutrino asymmetry. A final non-vanishing baryon asymmetry can then be obtained, provided the right-handed component remains out of equilibrium, while sphalerons are active. More specifically, successful sequestering requires interactions that would equilibrate the two chiral sectors to remain inefficient until after sphaleron freeze-out. This ordering allows a persistent nonzero baryon asymmetry to be generated before sphaleron processes cease. Subsequent left-right equilibration can redistribute the lepton asymmetry but cannot erase the previously generated BAU.

To better understand how the above can be realized in a given model, consider the collision term $\mathcal{C}$ of the Boltzmann equations describing the evolution of the leptonic asymmetry. It can be written schematically as
\begin{equation}\label{eq: general collision operator} 
    \mathcal{C}=\mathcal{S}\,+\,\mathcal{W}_{LR}\,+\,\mathcal{T}_{sph}\,,
\end{equation}
where $\mathcal{S}$ includes processes that source the chiral lepton asymmetry,
$\mathcal{W}_{LR}$ those that wash it out, and $\mathcal{T}_{sph}$ represents the lepton-to-baryon asymmetry conversion via EW sphalerons. In the present analysis, we do not concern ourselves with a specific asymmetry production mechanism, namely with concrete interactions behind $\mathcal{S}$ (see e.g. Refs.~\cite{Dick:1999je,Abel:2006hr,Berbig:2023uzs,Blazek:2024efd,Ahmed:2025vzl} for different production scenarios), but focus instead on the interplay between $\mathcal{W}_{\rm LR}$ and $\mathcal{T}_{\rm sph}$.

Importantly, each term in~\Cref{eq: general collision operator} is efficient over a finite temperature range. The weak sphaleron processes in the SM, inducing $\mathcal{T}_{sph}$, are active within a wide temperature range~\cite{Moore:1998zk,Bodeker:1999gx,Bento:2003jv,DOnofrio:2014rug}
\begin{equation}
   130\,\mathrm{GeV}\lesssim T\lesssim  10^{12}\,\mathrm{GeV}\,,
\end{equation}
while the source terms, contributing to $S$, on the other hand, are usually most efficient around a localized temperature, which we denote $T_S$. Finally, the efficiency of the left-right equilibration processes, sourcing $\mathcal{W}_{\rm LR}$, depends on the available operators communicating between the two chiral components. 

Crucially, when the efficiency ``windows'' of $\mathcal{W}_{LR}$ and $\mathcal{T}_{sph}$ overlap, the asymmetry generated by $\mathcal{S}$ will efficiently wash out. In the original Dirac leptogenesis scenario~\cite{Dick:1999je}, the smallness of the Dirac neutrino Yukawas guarantees that such an overlap does not occur, and therefore a non-vanishing baryon asymmetry can be generated. Indeed, a more general conclusion can be extracted from this work, i.e. given a renormalizable Yukawa term for a chiral fermion $f$ charged under $SU(2)_L$, with coupling $y_f$, the efficiency windows of $\mathcal{W}_{LR}$ and $\mathcal{T}_{sph}$ will not overlap during radiation domination provided that\footnote{We note that this relation assumes the sphaleron processes are as in the SM, but they could be modified in the presence of other gauge interactions.}~\cite{Dick:1999je,Besak:2012qm,Ghisoiu:2014ena,Bodeker:2014hqa}
\begin{equation}\label{eq: DL Yukawa bound}
    y_f<\mathrm{few}\,\times\,10^{-7}\,.
\end{equation}
If $\nu_R$ is introduced into the SM, such that a Yukawa interaction with the Higgs boson and the active neutrinos can generate light Dirac neutrino masses, Yukawa couplings of the order $y_\nu\sim\mathcal{O}(10^{-13})$ are needed to satisfy oscillation data. Since this value comfortably satisfies the bound in~\Cref{eq: DL Yukawa bound}, right-handed neutrinos are natural candidates to act as the necessary secluded reservoir needed for successful Dirac leptogenesis~\cite{Dick:1999je}. 

\begin{figure}
    \centering
    \includegraphics[width=0.85\linewidth]{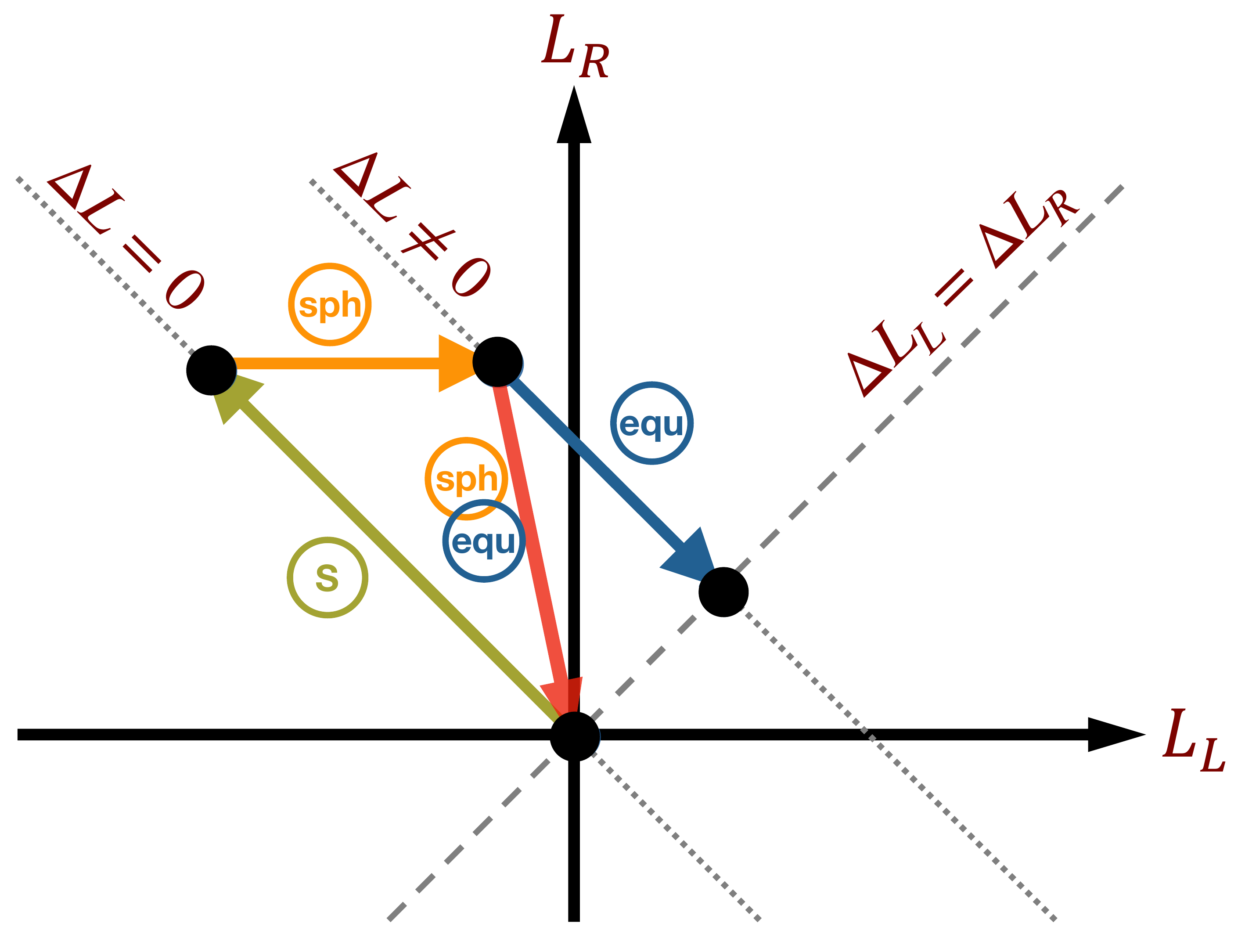}
    \caption{Schematic of the chiral lepton asymmetry evolution and relevant processes in Dirac leptogenesis. When left-right equilibration processes become efficient after sphalerons have frozen out (blue arrow), a BAU transferred from the leptonic sector is preserved until today. When their efficiency range overlaps with that of the sphalerons (red arrow), both lepton and baryon asymmetry are erased. See main text for more details. 
    }
    \label{fig: DL diagram}
\end{figure}

The diagram in~\Cref{fig: DL diagram} summarizes the mechanism described above. Starting from symmetric initial conditions, where the lepton number stored in left- and right-handed fermions, denoted by $L_L$ and $L_R$, respectively, is zero, a source term becomes efficient and leads to a period of asymmetry production ($\text{s}$) that generates a non-vanishing asymmetry in the chiral components while global lepton number remains conserved ($\Delta L = \Delta L_L+\Delta L_R=0$). Weak sphalerons ($\text{sph}$), while active, partially convert the left-handed lepton asymmetry into a baryon asymmetry. Next, there are two possibilities. If the left-right equilibration processes ($\text{equ}$) become efficient only after sphalerons have shut off, they drive the system to $L_L = L_R$ without changing the previously generated baryon asymmetry. This is represented in the diagram with a blue arrow, and can be realized for Dirac neutrinos as shown in Ref.~\cite{Dick:1999je}. On the other hand, if there is a temperature range below the asymmetry production scale for which ($\text{sph}$) and ($\text{equ}$) become efficient simultaneously, the generated asymmetry will be fully washed out in both the baryonic and leptonic sectors. This is represented by the red arrow and is realized by any fermion not satisfying the bound in~\Cref{eq: DL Yukawa bound}, which is the case for all charged fermions in the SM. The precise washout trajectory depends on the interplay of sphalerons and equilibration processes, and is, for simplicity, depicted as a straight line. Clearly, in the case when the equilibration becomes efficient before sphalerons, there will no longer be a lepton asymmetry to convert, also preventing a successful BAU.

Note that~\Cref{eq: DL Yukawa bound} assumes renormalizable Yukawa couplings during radiation domination. Under these assumptions, washout effects become efficient \textit{below} a certain temperature $T_{\rm LR}^{\rm{dim=4}}$. However, if the Yukawa interactions are replaced by higher-dimensional operators involving additional fields, the corresponding washout processes become efficient \textit{above} a given temperature $T_{\rm LR}^{\rm{dim>4}}$.\footnote{This statement is valid only while the EFT is valid, since at large enough temperatures the EFT breaks down and the effective interactions must be replaced by the fundamental ones.} The dimension-five Weinberg operator exemplifies this behavior, since the $\Delta L\neq0$ washout processes it mediates, which are relevant for Majorana leptogenesis, are efficient only at high-enough temperatures. 

This temperature dependence suggests a way to extend asymmetry sequestering to SM charged leptons, whose renormalizable Yukawa couplings do not satisfy~\Cref{eq: DL Yukawa bound}. If their mass-generating interactions remain higher-dimensional while sphalerons are active, left-right equilibration can become inefficient as the Universe cools, opening a temperature window in which a chiral asymmetry can be generated and partially converted into a baryon asymmetry. The observed Yukawa couplings may then emerge after sphaleron freeze-out, when subsequent left-right equilibration can no longer erase the BAU. In the next section, we describe how this idea can be realized within a Froggatt-Nielsen framework~\cite{Froggatt:1978nt} whose flavor symmetry remains unbroken until after sphaleron freeze-out. We examine the conditions under which the higher-dimensional interactions decouple sufficiently early for SM right-handed charged leptons to serve as asymmetry reservoirs, while reproducing their observed masses after FN-symmetry breaking.
%
\section{Dirac leptogenesis in a Froggatt-Nielsen construction}\label{sec:DLinFN}

\subsection{Setup}
\begin{figure}[t!]
    \centering
    \includegraphics[width=0.85\linewidth]{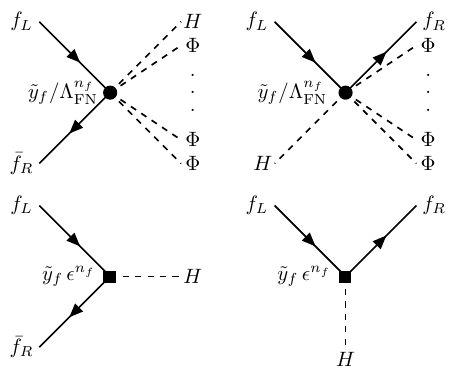}
    \caption{Effective diagrams contributing to left-right equilibration of the Dirac fermion $f$. The dots in the upper diagrams signify $n_f$ flavons in the final state.}
    \label{fig:FN diagrams}
\end{figure}

In an FN realization of the SM, the fermionic mass-generating operators are, except for the heaviest fermions, higher-dimensional~\cite{Froggatt:1978nt}. More specifically, the chiral SM fermions carry different $\mathcal{O}(1)$ charges under a global $U(1)_{\mathrm{FN}}$ symmetry, and there exists a complex scalar SM singlet $\Phi$, the flavon, with a non-zero FN charge, e.g. $q_\Phi^{FN}=-1$, while the Higgs is uncharged.
After symmetry breaking, once $\langle \Phi\rangle = v_\Phi/\sqrt{2}$, the complex flavon field can be written in terms of an axial Nambu-Goldstone mode $a$, called Axion-Like Particle (ALP) in the following, and a radial mode $\rho$ as 
\begin{equation}
    \Phi = \frac{v_\Phi + \rho}{\sqrt{2}} e^{i a / v_\Phi} \, .
\end{equation}
As a consequence of the $U(1)_{\rm FN}$ symmetry, the invariant Yukawa interactions now contain, in addition to the Higgs, $n_f\geq 0$ insertions of the new field $\Phi$. For a generic fermion $f$, with left-handed component contained in an $SU(2)_L$ doublet $\Psi$, the fermionic masses come from interactions of the type
\begin{equation}\label{eq: FN eff Lagrangian}
    -\mathcal{L}^{\mathrm{eff}}_{FN}\supset c_f^{ij}\,\Big(\frac{\Phi}{\Lambda_{\mathrm{FN}}}\Big)^{n_f^{ij}}\,\overline{\Psi}_{L,\,i}\,H_f\,f_{R,\,j}\,+\,\mathrm{h.c.}\,,
\end{equation}
with $H_f=H$ for down quarks and charged leptons, and $H_f=i\,\sigma_2\,H^*$ for up quarks and neutrinos, where $H$ is the Higgs doublet. These interactions generate the diagrams shown in the upper panel of~\Cref{fig:FN diagrams}. The coefficients $c_f^{ij}$ are of $\mathcal{O}(1)$, since they are a function of the ultraviolet couplings that are assumed to be of $\mathcal{O}(1)$. The exponent $n_f^{ij}$ depends on the charge assignments of the fermions under the $U(1)_{\mathrm{FN}}$ symmetry, and the indices $\{i,j\}$ denote different flavors. Such terms can originate e.g. from integrating out vector-like fermion mediators with masses around the scale $\Lambda_{\mathrm{FN}}$. Finally, the operator with $n_f$ flavons has a mass dimension of $4+n_f$, rendering it higher-dimensional for $n_f>0$.

We call $T_{\mathrm{FN}}$ the temperature of the Universe at which the Froggatt-Nielsen phase transition (FNPT) occurs and $U(1)_{\mathrm{FN}}$ is broken spontaneously, such that the flavon gets a vacuum expectation value (VEV), $\langle\Phi\rangle\equiv v_\Phi/\sqrt{2}$. Once this happens, effective renormalizable Yukawa couplings arise, namely\footnote{For the leptogenesis part, we ignore flavor indices. 
}
\begin{equation}\label{eq: 4-d Yukawa}
    -\mathcal{L}_{\mathrm{Yuk}}^{FN}=y_{\mathrm{eff}}^f\,\overline{\Psi}_L\,H_f\,f_R\,+\,\mathrm{h.c.}\quad \mathrm{for}\quad T<T_{\mathrm{FN}}\,,
\end{equation}
with the effective Yukawa $y_{\mathrm{eff}}^f$ given in terms of the parameter $\epsilon\equiv v_\Phi/(\sqrt{2}\Lambda_{\mathrm{FN}})$ as
\begin{equation}
    y_{\mathrm{eff}}^f=c_f\,\epsilon^{n_f}\,. 
\end{equation}
These effective interactions give rise to the diagrams shown in the lower panel of~\Cref{fig:FN diagrams}. A small hierarchy $v_\Phi < \Lambda_{\mathrm{FN}}$ then naturally leads to hierarchical fermion masses, since below the electroweak phase transition (EWPT) temperature $T_{\rm EW}$ and below $T_{\rm FN}$,~\Cref{eq: 4-d Yukawa} leads to
\begin{equation}
    m_f=c_f\,\epsilon^{n_f}\,\frac{v}{\sqrt2}\,,
\end{equation}
with $v$ the Higgs VEV. Lighter fermions require larger values of $n_f$, since it leads to larger powers of the suppression factor $\epsilon$. With the Cabibbo angle $\sim0.23$, we assume the expansion parameter to be of the same order of magnitude, namely $\epsilon \sim \mathcal{O}(0.1)$, in the following. If a different $\epsilon$ were chosen, the $c_f$ values would have to be tuned in order to obtain the correct fermion masses and mixings.
\subsection{Avoiding washout}
The effective interactions in~\Cref{eq: FN eff Lagrangian,eq: 4-d Yukawa} generate annihilation and scattering processes that contribute to the chiral-lepton-asymmetry washout $\mathcal{W}_{LR}$. Representative diagrams are shown in~\Cref{fig:FN diagrams}, where the squared effective vertex is only valid below $T_{\rm FN}$. Further relevant tree-level diagrams can be obtained by attaching, e.g., a top-quark pair or soft gauge bosons to the Higgs leg. Even though these extra diagrams can contribute with the same order of magnitude as the ones considered in~\Cref{fig:FN diagrams}, we ignore them in our calculations as the qualitative conclusions are expected to remain robust under their inclusion. For more details on these additional effects, we refer the reader to Refs.~\cite{Besak:2012qm,Ghisoiu:2014ena,Bodeker:2014hqa}, and to~\appsref{sec: th mass effects}{sec: beyond MB} where the effects of thermal masses and boson enhancements are estimated, respectively.

To determine whether or not the left-right equilibration processes related to fermion $f$ are efficient at a given temperature $T$, we must compute the corresponding thermally-averaged interaction rate $\langle\Gamma_{\mathrm{LR}}^f\rangle$. For the tree-level left-right annihilation rate corresponding to the diagram in the top-left corner of~\Cref{fig:FN diagrams} and in the limit where the center-of-mass energy $s$ is much larger than the masses of the involved particles, $s\gg\{m_i\}$, the rate is given by (see~\appref{sec: app LR annihilation unb} for more details)
\begin{equation}
    \langle\Gamma_{\mathrm{LR}}^f\rangle(T)=\kappa_f^{d>4}\,|c_f|^2\,\frac{T^{2n_f+1}}{\Lambda_{\mathrm{FN}}^{2n_f}}\,,
\end{equation}
with
\begin{equation}
    \kappa_f^{d>4}=\frac{n_f\,(n_f+1)!}{8\,(2\pi)^{2n_f+1}}\,.
\end{equation}
\begin{figure*}[t!]
    \centering
    \includegraphics[width=0.99\linewidth]{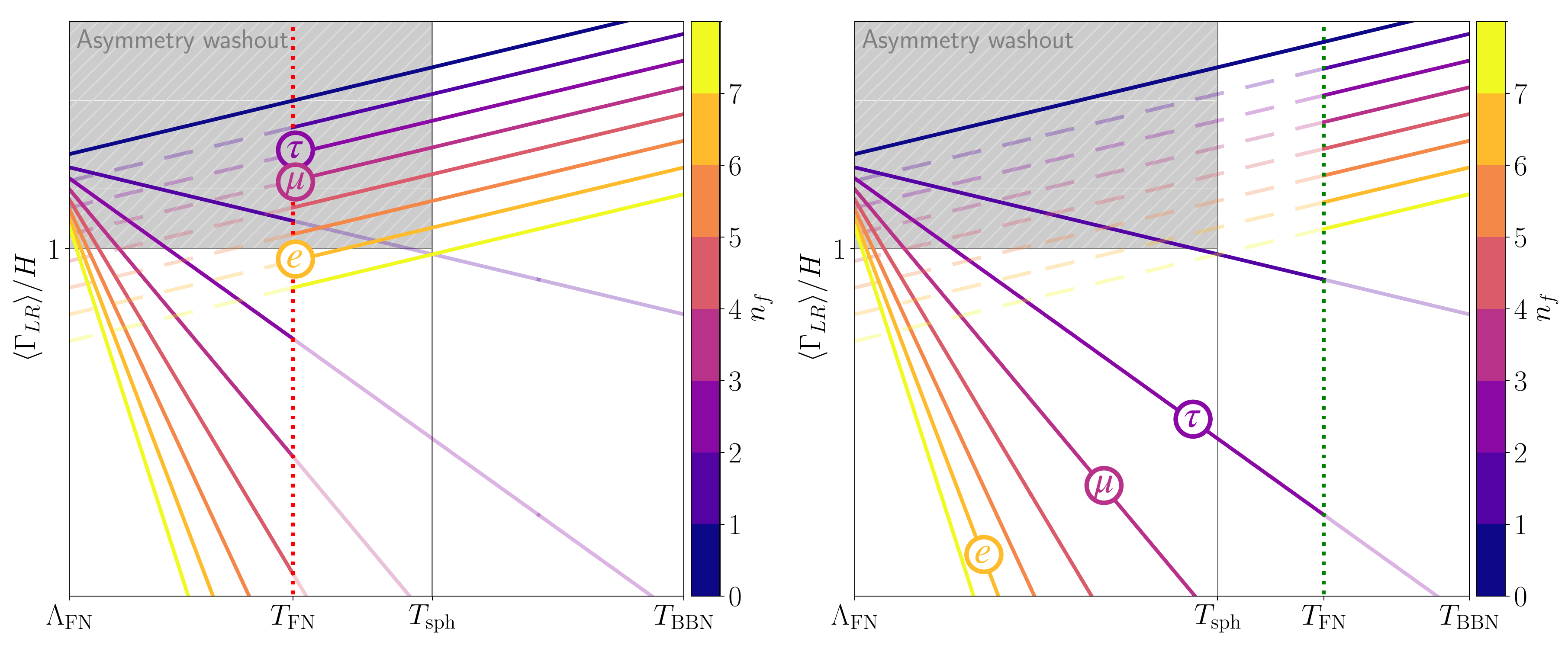}
    \caption{Illustration of the scaling of the ratio $\langle\Gamma_{LR}\rangle/H$ (solid lines) for $c_f\sim\mathcal{O}(1)$, $\epsilon=0.1$, and for different values of $n_f$, i.e. different numbers of flavons in the operator of~\Cref{eq: FN eff Lagrangian}, which generates the hierarchical fermion masses. We label the curves reproducing the observed charged lepton Yukawas in a way that is compatible with the FN charge assignments specified in~\Cref{tab: FN eps=0.1}. The shaded region corresponds to temperatures where the efficiency windows of $\mathcal{W}_{LR}$ and $\mathcal{T}_{sph}$ overlap. The dashed lines depict the equilibration via effective four-dimensional interactions, which for $n_f>0$ are only possible below the temperature at which the FN symmetry is broken (dotted vertical line), and represented by the dashed line becoming solid. There, the interaction including external flavons, which has the opposite temperature scaling, becomes subleading, as illustrated by the corresponding lines becoming transparent.  \textit{Left:} When the FNPT happens before sphalerons freeze out, equilibration becomes efficient simultaneously with sphaleron processes, leading to a washout of a previously stored asymmetry, unless $n_f$ is large enough. \textit{Right:} When the FNPT is delayed until after the sphalerons have frozen out, the equilibration of the chiral lepton asymmetry can no longer be communicated to the baryon sector. Thus, a BAU can remain until today, provided also that the initial asymmetry was injected sufficiently late, i.e. at temperatures after the solid lines have left the gray washout region; see \Cref{sec:AllowedProduction} for details.
    }
    \label{fig: GLR/H scaling general n}
\end{figure*}

When the flavon obtains a non-vanishing VEV, effective 4-dimensional Yukawas, given by~\Cref{eq: 4-d Yukawa} and represented by the square vertices in~\Cref{fig:FN diagrams}, contribute to the left-right equilibration rate. For the bottom left diagram in~\Cref{fig:FN diagrams}, one obtains (see~\appref{sec: app LR annihilation brok} for more details)
\begin{equation}\label{eq: Gamma LR 4-dim}
    \langle\Gamma_{\mathrm{LR}}\rangle(T)\simeq \kappa_f^{d=4}\,|y_{\mathrm{eff}}^f|^2\,T\,,
\end{equation}
with $\kappa_f^{d=4}=\mathrm{few}\times10^{-3}$. In particular, for the annihilation process $f_L+\bar f_R\to t+\bar t$ and $f_L+\bar f_R\to H+V_\mu$, as well as their scattering counterparts, one obtains $\langle\Gamma_{\mathrm{LR}}\rangle\sim y_t^2\,|y_{\mathrm{eff}}^f|^2\,T$ and $\langle\Gamma_{\mathrm{LR}}\rangle\sim g^2\,|y_{\mathrm{eff}}^f|^2\,T$, respectively, with $y_t$ the top Yukawa and $g$ a gauge coupling. Clearly, the result in~\Cref{eq: Gamma LR 4-dim} also holds for heavy fermions with $n_f=0$ with the effective Yukawa replaced by the fundamental one.

Processes mediating left-right equilibration become efficient within temperatures for which
\begin{equation}
    \langle\Gamma_{LR}\rangle(T)\gtrsim H(T)\,,
\end{equation}
where $H(T)$ is the Hubble rate, which in the radiation-dominated\footnote{We are primarily interested in the period in which sphalerons are active, i.e. before the EWPT, when the SM energy density scales as radiation in standard settings. In scenarios where this does not hold, e.g. in the presence of a period of matter or vacuum domination before the EWPT, the Hubble rate decreases more slowly with temperature relative to the radiation-dominated case, and as a result the equilibration becomes efficient later in the $d=4$ case, which is utilized in Ref.~\cite{Chen:2019wnk} to allow electrons to sequester the asymmetry. Thus, assuming a radiation-dominated era prior to the EWPT is a conservative choice in this work.} era reads
\begin{equation}\label{eq: Hubble rate}
    H(T)=\sqrt{\frac{8\pi^3\,g_*}{90}}\,\frac{T^2}{M_{\mathrm{Pl}}}\,,
\end{equation}
with $g_*$ the effective number of relativistic degrees of freedom of the thermal plasma and $M_{\mathrm{Pl}}$ the reduced Planck mass. 

Based on the rates derived above, we can determine the efficiency windows for the left-right equilibration processes of a fermion $f$ that couples to $n_f$ flavons. To that end, we define $T_{\rm LR}$ as the temperature at which the equilibration rate equals the Hubble rate, which can take different values depending on the FN charge assignments and whether or not the $U(1)_{\rm{FN}}$ symmetry is broken.

First, for $n_f=0$, i.e. as for the SM Yukawas, the temperature range at which the equilibration rate surpasses the Hubble rate is given by
\begin{equation}\label{eq: efficiency windows d=4}
        T<T_{\rm{LR}}=\sqrt{\frac{90}{8\pi^3 g_*}}\,\kappa_f^{d=4}\,|c_f|^2\,M_{\mathrm{Pl}}\, . 
\end{equation}
Second, with $n_f>0$, and for temperatures below $\Lambda_{\rm{FN}}$ the corresponding range reads
\begin{equation}\label{eq: efficiency windows d>4}
    T>T_\textrm{LR}=\Lambda_{\mathrm{FN}}\left(\frac{\Lambda_{\mathrm{FN}}}{M_{\mathrm{Pl}}}\frac{1}{\kappa_f^{d>4}\,|c_f|^2}\sqrt{\frac{8\pi^3 g_*}{90}}\right)^{\frac{1}{2n_f-1}} \, , 
\end{equation}
before the FNPT, and
\begin{equation} \label{eq: efficiency windows EFT}
    T<T_\textrm{LR}=\sqrt{\frac{90}{8\pi^3 g_*}}\,\kappa_f^{ d=4}\,|y_{\mathrm{eff}}^f|^2\,M_{\mathrm{Pl}} \,  
\end{equation}
after the FNPT.

It becomes evident from the lower bound on $T_\textrm{LR}$ in \Cref{eq: efficiency windows d>4} that the washout rate associated with fermion $f$ decreases for falling temperatures in the range $T_{\rm FN}<T_\mathrm{UV} < \Lambda_\mathrm{FN}$ when $n_f>0$. This is due to the fact that, in the higher-order nature of the interactions, the ratio $\Phi/\Lambda_\mathrm{FN}$, which is set to the fixed value of $v_\Phi / \Lambda_\mathrm{FN} \sim 0.1$ only after the flavon acquires a nonzero VEV, appears as $m_\mathrm{therm}/\Lambda_\mathrm{FN} \sim T / \Lambda_\mathrm{FN}$, where $m_{\rm th}$ corresponds to the flavon thermal mass. The flavons have to be present as physical fields in the equilibration processes, and their thermal masses decrease as the Universe cools down. Then, as temperature decreases, the ratio decreases more and more, further suppressing the equilibration rate. After symmetry breaking, the same temperature scaling as in the SM is recovered by setting the flavon to its VEV, which gives a contribution to the equilibration rate that dominates over the processes with dynamical flavons when $T \ll \Lambda_\mathrm{FN}$. 

The derived scaling of the ratio $\langle\Gamma_{LR}\rangle/H$ for different fermions is shown in~\Cref{fig: GLR/H scaling general n}, for $\epsilon=0.1$. For fundamental or effective $d=4$ Yukawas, the ratio is monotonically increasing with temperature and firmly inside the wash-out region for electron, muon, and tau leptons (see \Cref{tab: FN eps=0.1}). 
On the other hand, for $n_f>0$ and temperatures above the FN breaking scale, the ratio decreases with temperature, and for low-enough values of $T_{\rm FN}$ relative to $\Lambda_{\rm FN}$ it exits the shaded wash-out region where $\mathcal{W}_{\rm LR}$ and $\mathcal{T}_{\rm sph}$ overlap. In that regime, a generated asymmetry is safe from washout. However, as soon as the flavon acquires a non-vanishing VEV, the renormalizable effective vertices will also contribute to the equilibration of the chiral asymmetry and will generally dominate over the processes with $n_f$ flavons in the final state. We note that the number of decreasing lines that are able to exit the washout region before sphaleron freezeout is a function of $\Lambda_{FN}$, which becomes clear in~\Cref{fig: T_LR Lambda_F}. For this illustration we choose a large-enough FN scale, $\Lambda_{FN}=5\times 10^8\,\rm{GeV}$, so that all fermions with $n_f\geq1$ can technically constitute viable reservoirs.

Crucially, if $T_{\rm FN}>T_{\rm EW}$ and $\epsilon=0.1$, the left-right equilibration processes associated with fermions with $n_f<7$, which would include all fermions present in the SM other than neutrinos, become efficient before sphalerons freeze out, occurring at $T_{\rm sph}\approx T_{\rm EW}$. As a consequence, any previously generated asymmetry within these fermions is efficiently washed out, as represented in the left plot of~\Cref{fig: GLR/H scaling general n} by the corresponding lines entering the shaded region before $T_{\rm sph}$. Thus, for a baryon asymmetry originating from a leptonic asymmetry within SM charged leptons to survive until today, the FN symmetry breaking has to be delayed until after the sphalerons have shut off. The right plot of~\Cref{fig: GLR/H scaling general n} depicts such a scenario.

This requirement together with constraints on the flavon VEV from flavor observables, which will be detailed in the next section, results in the necessity of a hierarchy of scales
\begin{equation}
    v_\Phi\gg T_{\rm FN}\,,
\end{equation}
that can be realized via a supercooled FN phase transition.

To sum up, these results show that for dimension-four Yukawas there is an upper bound on the Yukawa coupling in order to protect the chiral asymmetry from being washed out, which follows from requiring that the efficiency window of $\mathcal{W}_{LR}$ must not overlap with that of $\mathcal{T}_{sph}$. On the other hand, when the Yukawa interactions remain higher-dimensional until after $T_{\rm sph}$, the chiral asymmetry washout dynamics is significantly modified and a completely different scale setting of Dirac leptogenesis emerges. Most importantly, and as will be shown in detail in~\Cref{sec:AllowedProduction}, this alternative scenario allows other leptons within or beyond the SM to successfully mediate leptogenesis. In particular, SM charged leptons, namely electrons, muons, and tau leptons, can, under this framework, constitute viable asymmetry-sequestering reservoirs.
\subsection{Allowed production scales}\label{sec:AllowedProduction}
Even though we remain agnostic to the asymmetry production mechanism, the above results allow us to infer the range of temperatures within which such a mechanism would need to become efficient in order to successfully generate the BAU.
\begin{figure}[!t]
    \centering
    \includegraphics[width=0.99\linewidth]{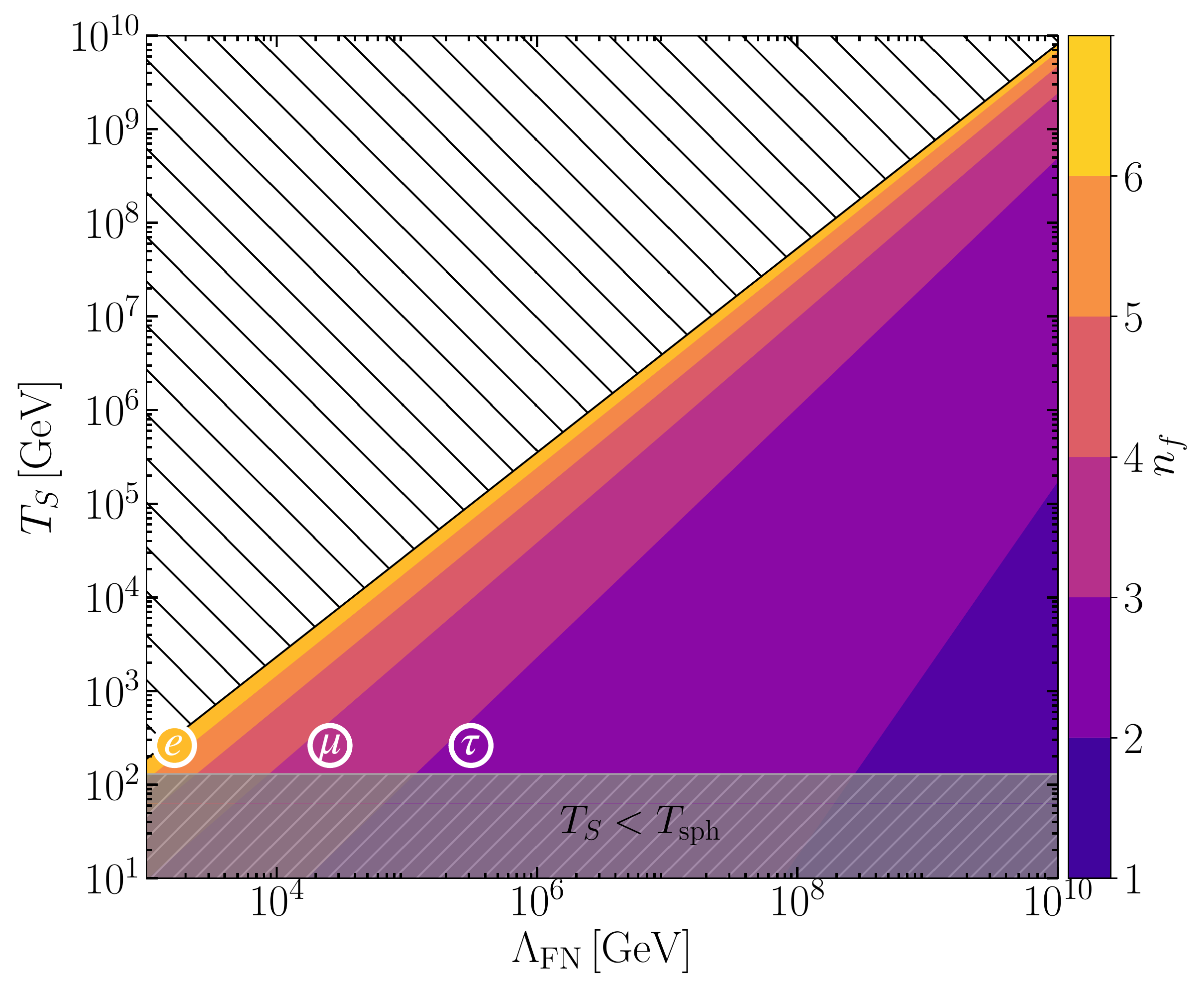}
    \caption{Allowed values of the asymmetry injection temperature for different values of $n_f$ and as a function of the EFT scale $\Lambda_{\mathrm{FN}}$, for $T_{\rm FN}< T_{\rm sph}$.}
    \label{fig: T_LR Lambda_F}
\end{figure}

The conditions that must be satisfied by the asymmetry production scale $T_S$, the equilibration scale $T_{\rm LR}$, and the FN breaking scale $T_{\rm FN}$ in order to avoid washout before the sphaleron freeze-out scale $T_{\rm sph}$ can be summarized as
\begin{equation}
    \begin{cases}
    T_S>T_{\rm sph}>T_{\rm LR}\quad&\mathrm{if}\quad n_f=0\,,\\
    T_{\rm LR}>T_S>T_{\rm sph}>T_{\rm FN}\quad&\mathrm{if}\quad 0<n_f<n_f^c\,,\\
    T_{\rm LR}>T_S>T_{\rm sph}\quad&\mathrm{if}\quad n_f \geq n_f^c\,,
    \end{cases}
\end{equation}
where the $T_\textrm{LR}$ are defined in \Cref{eq: efficiency windows d=4,eq: efficiency windows d>4,eq: efficiency windows EFT} and $n_f^c=7$ for $\epsilon=0.1$ and $c_f\sim {\cal O}(1)$. 

In~\Cref{fig: T_LR Lambda_F}, the colored regions show the temperature scales $T_S$ at which the asymmetry production would need to take place in order to obtain a non-vanishing BAU today, as a function of the EFT scale $\Lambda_{\rm FN}$. 
It is important to note that requiring that the source becomes active before the sphaleron freeze-out temperature corresponds to a necessary but not sufficient condition. Realistically, the process of asymmetry injection plus sphaleron conversion is not an instantaneous one, and therefore the source should become efficient early enough for the overall process to be completed before sphalerons shut off.

The region spanned by the allowed values increases with $n_f$. The case $n_f=0$ is not included given that the associated rate is independent of $\Lambda_{\rm FN}$. Exponents satisfying $n_f\geq n_f^c$ are not included simply because the corresponding Yukawa couplings can trivially sequester the asymmetry even in their renormalizable form, as in the case of neutrinos, and the injection temperature can be nearly as large as the UV scale. The smaller $n_f$ is, the higher $\Lambda_\textrm{FN}$ has to be so that $T_\textrm{LR}$ (and therefore $T_S$) can be larger than $T_\textrm{EW}$, after which an injected asymmetry could no longer be transferred to the baryon sector via sphalerons. In \Cref{fig: GLR/H scaling general n}, this requires that the lines exit the shaded region before crossing $T_\textrm{EW}$. For instance, if $\Lambda_\textrm{FN}$ were brought slightly closer to $T_\textrm{EW}$, this would no longer be the case for the $n_f = 1$ line.
\begin{table}[!t]
    \centering
    \begin{tabular}{|c|c|c|c|c|c|}
    \hline
        $f$ & top & charm & up & tau & electron\\
        \hline
         $n_f$ & 0 & 2& 5& 2& 6\\
         \hline
         $f$ & bottom & strange & down & muon & neutrino\\
         \hline
         $n_f$ & 2& 3& 5& 3& 12 \\
         \hline
    \end{tabular}
    \caption{ Exemplary values of $n_f = n_{f_L} + n_{\bar{f}_R}$ for the different fermions in the SM, with $\epsilon=0.1$.}
    \label{tab: FN eps=0.1}
\end{table}

For concreteness, consider an explicit FN realization of the SM by fixing $\epsilon=0.1$ and $c_f\sim\mathcal{O}(1)$. 
\Cref{tab: FN eps=0.1} shows the values of $n_f$, which are determined by the sum of the charges of the chiral fields, i.e. $n_f \equiv n_{f_L} + n_{\bar{f}_R}$,\footnote{Here, we assume diagonal Yukawas for simplicity, but we consider specific textures in \Cref{sec:flavObs}. Qualitatively, the results remain unchanged for realistic values.} that can reproduce the hierarchical SM Yukawas when $\epsilon \sim 0.1$. We note that when varying the $\mathcal{O}(1)$ numbers $c_f$, the $n_f$ in \Cref{tab: FN eps=0.1} can differ slightly. 

From \Cref{fig: T_LR Lambda_F} it is then possible to read off a lower bound on $\Lambda_{\rm{FN}}$ for the SM leptons. Specifically, for the tau lepton ($n_f = 2$) there is a non-vanishing temperature range in which the asymmetry can be injected and remain successfully sequestered if $\Lambda_\textrm{FN}>10^5\,\text{GeV}$, while for the muon ($n_f = 3$) $\Lambda_\textrm{FN}>10^4\,\text{GeV}$ is required, and for the electron ($n_f = 6$), we find that $\Lambda_\textrm{FN}$ can be as low as $\mathcal{O}(10^3)\,\text{GeV}$. 
Note, however, that $n_f=6$ is still smaller than $n_f^c=7$, which means that if the FN breaking scale is larger than the sphaleron shut-off scale, the corresponding left-right equilibration rate will become efficient simultaneously with the sphalerons and any previously generated electron asymmetry will be washed out. 
Finally, for Dirac neutrinos one obtains $n_f=12>n_f^c$, and they can comfortably store an asymmetry regardless of whether or not the FN breaking happens before or after the EWPT. Although interesting, this is not surprising given that Dirac neutrinos are good asymmetry reservoirs even with fundamental Yukawas. 
%
\section{Experimental Froggatt-Nielsen Bounds}\label{sec:flavObs}
A universal prediction in Froggatt-Nielsen models are off-diagonal Yukawa couplings between the flavon and fermions, which introduce flavor violation into the model. Due to the non-observation of such processes, the FN scale, identified with the masses of the vector-like fermions mediating the Yukawa interactions, has to be much larger than the EW scale.
When both the axial and radial modes are sufficiently heavy, the dominant constraint comes from integrating out the heavy messenger fermions, resulting in a universal bound on $\Lambda_\mathrm{FN}$. However, when a light scalar mode exist, the loop diagram involving it can dominate and lead to more stringent constraints on $\Lambda_\mathrm{FN}$.

The axial mode can obtain its mass from some explicit symmetry breaking, rendering it a pseudo-Nambu-Goldstone boson.  If its potential is exclusively generated by a non-vanishing QCD anomaly of the global $U(1)_{\rm FN}$ symmetry, it can solve the strong CP problem, connecting its resolution to that of the flavor puzzle~\cite{Calibbi:2016hwq,Ema:2016ops,Ema:2018abj,Alanne:2018fns}, and potentially explain the Dark Matter abundance in regions of the parameter space.
However, for the sake of this analysis, we treat its mass as a free parameter, so it can be light or heavy. On the other hand, as will be described in detail in \Cref{sec:PT}, the mass of the radial mode $m_\rho$ is bounded from above by requiring the delay of the FNPT until after EW sphalerons freeze out. This follows from a supercooled potential needing to be flat enough, i.e. the curvature at the VEV must be small. Therefore, we expect at least the radial mode to give an unavoidable contribution to the flavor-violating processes, and the constraint on $\Lambda_\mathrm{FN}$ to carry a mass dependence, in the region of interest.

In the following, we focus on the radial mode contribution and highlight the dominant constraints for an FN symmetry realized only in the leptonic sector, which corresponds to the minimum requirement for the proposed leptogenesis mechanism to work. We also comment on the additional constraints that must be included when generalizing the FN to include all fermions, which would be needed to explain all flavor hierarchies. We relegate the corresponding details to~\appref{app:flavorbounds}.
\subsection{Constraints from lepton flavor violation}
There are strong bounds on $v_\Phi$ from rare lepton decays \cite{Jodidio:1986mz,TWIST:2014ymv,Belle-II:2022heu} when $\tau \rightarrow \mu +\rho$ and $\mu \rightarrow e +\rho$ are kinematically allowed, i.e. when $m_\rho \leq (m_\tau - m_\mu) = 1.67\,\textrm{GeV}$ for the former case. We set this as a lower bound for the flavon mass and keep in mind that it might be possible to construct a model in which these constraints are alleviated to some extent. For flavons heavier than the SM leptons, the most stringent constraint comes from charged lepton flavor violation (CLFV), especially from the radiative decay $\mu \rightarrow e \gamma$, which is constrained by the MEG II experiment to be \cite{MEGII:2025gzr}
\begin{equation}
    \text{BR}(\mu \rightarrow e \gamma) < 1.5 \times 10^{-13} \, .
    \label{eq:muegammabound}
\end{equation}
The branching ratio is given by \cite{Greljo:2024evt}
\begin{equation}
    \rm{Br} (\mu \rightarrow e \gamma) = \frac{m_\mu^3}{4 \pi \Gamma_\mu} \left( \left| C_{e\gamma}^{\mu e} \right|^2 +\left| C_{e\gamma}^{e \mu} \right|^2 \right)
\end{equation}
with
\begin{align}
    C_{e\gamma}^{ij} = & \frac{e}{32 \pi^2 m_{\rho}^2}   \sum_{k=1,2,3} \Big( \frac{1}{6} \left( m_{l_j} c_{ki}^{e\ast} c_{kj}^e + m_{l_i} c_{ik}^e c_{jk}^{e\ast}\right) \nonumber \\ 
    &+ m_{l_k} c_{ki}^{e\ast} c_{jk}^{e\ast} \left( 3/2 + \log m_{l_j}^2/m_{\rho}^2 
    \right) \Big) \, .
    \label{eq:Cij_e}
\end{align}
The couplings $c_{ij}$ are derived in~\appref{app:flavorbounds}. 
Lepton masses and the muon decay width are taken from Ref.~\cite{ParticleDataGroup:2026mpi}.

\begin{figure}
    \centering
    \includegraphics[width=\linewidth]{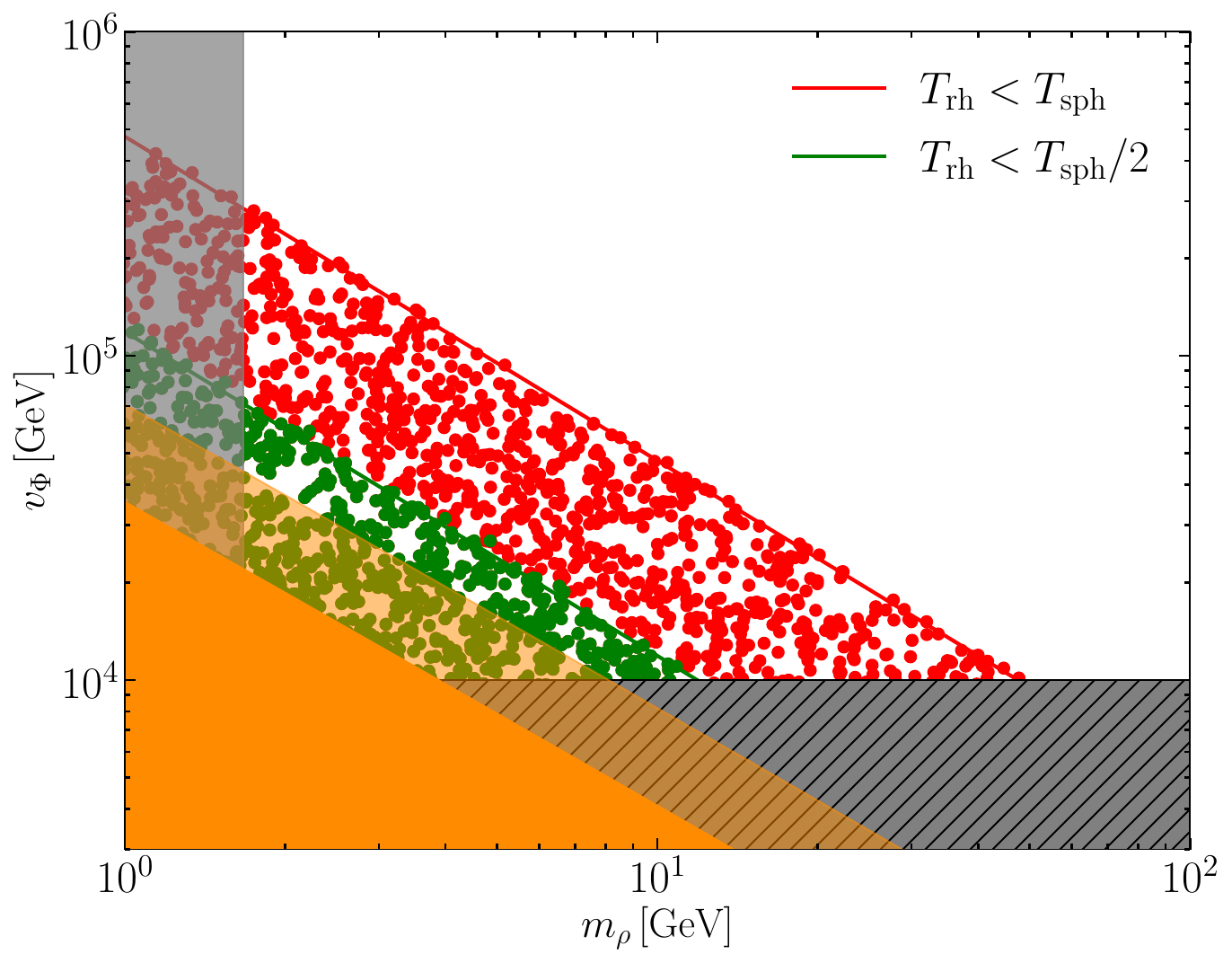}
    \caption{CLFV bounds from $\mu \rightarrow e \gamma$. For a flavon running in the loop, the naive constraint is shown in light orange, and the one with texture suppression in orange. The contribution from the VL messenger fermions constrains $v_\Phi>10^4\,\mathrm{GeV}$ (grey lower region), which dominates for large $m_\rho$. Masses below $1.67\,\mathrm{GeV}$ are excluded due to rare lepton decays (left grey region). The origin of the scatter points and lines are explained in detail in~\Cref{sec:PT}.
    }
    \label{fig:Leptonic sector}
\end{figure}

\Cref{fig:Leptonic sector} shows the allowed parameter space of the model. There are two main constraints. Firstly, an upper bound (solid red line) on $m_\rho$ and $v_\Phi$ that results from requiring that the FNPT does not reheat the Universe above $T_{\rm sph}$, as it would re-activate the sphalerons while keeping the $U(1)_{\rm FN}$ symmetry broken,\footnote{This follows from the fact that the reheating temperature can never be larger than the critical temperature.} which would lead to a washout of the previously generated asymmetry. The colored points result from a numerical scan, described in detail in~\Cref{sec:PT}, of the input parameters of the model.

Secondly, a lower bound on the flavon VEV and mass from lepton flavor violation. Due to the rare lepton decays we consider $m_\rho > 1.7\,\textrm{GeV}$ (transparent grey region), while for larger masses the dominant bound comes from the non-observation of $\mu \rightarrow e \gamma$ (\Cref{eq:muegammabound}), shown in light orange. Even though the constraints are given in terms of the mass of the radial mode, the ALP contribution can be obtained by replacing $c \rightarrow i c$; see~\Cref{eq:FCinteractions}. In the case where the ALP is much lighter than the radial mode, it gives the leading constraints on $v_\Phi$. If they have similar masses, their contributions to the branching ratio cancel to some extent. We describe in more detail how the bounds change depending on the ALP mass in~\appref{app:flavorbounds}. Here, we treat the mass of the ALP as a free parameter, and thus focus only on the constraints coming from the radial mode, keeping in mind that in a specific model the constraints may become less or more stringent.

Flavor violation originates from off-diagonal components of the Yukawa matrix, however in our leptogenesis analysis we focus on the diagonal Yukawas, which leaves some freedom in choosing the individual $U(1)_{\rm FN}$ charge assignment for the left- and right-handed fermion fields.
This gives rise to different ``textures'' of the Yukawa matrices. As shown in Ref.~\cite{Cornella:2024jaw}, specific textures can decrease the bound on $\Lambda_\mathrm{FN}$ by half an order of magnitude. This possibility is shown in~\Cref{fig:Leptonic sector} with a darker orange shading. We note that the precise bounds determined in Refs.~\cite{Cornella:2024jaw,Cornella:2023zme} cannot be directly applied to our case, since their work is based on SMEFT, while we allow for a light flavon. Therefore, we merely adopt the relative changes determined there to our numerical benchmark. We do, however, take their constraints as a lower bound on $v_\Phi$, which becomes relevant for larger flavon masses, where the contribution from the integrated-out messenger fermions dominates. Taking into account the different normalizations, we set ${v_\Phi}_{\rm min} = 10^4\,\textrm{GeV}$, shown as the lower gray region in~\Cref{fig:Leptonic sector}. 
\subsection{FN in the leptonic and quark sectors}
For charged leptons to be able to successfully sequester a chiral asymmetry following the prescription proposed in this work, it is necessary and sufficient to charge only these fermions under $U(1)_\textrm{FN}$. Thus, we treat the constraints from $\mu \rightarrow e \gamma$ as unavoidable. Ideally, however, quarks should also carry $U(1)_{\rm FN}$ charges, so that all fermion hierarchies can be explained. In this case, additional bounds must be considered and, in particular, regions with $m_s <m_K, m_B$ are very constrained due to invisible meson decays \cite{Greljo:2024evt}. This then translates to the necessity of many orders of magnitude in supercooling, i.e. $T_\mathrm{FN}\lesssim T_\textrm{EW}\lll v_\Phi$, which is incompatible with not reheating the Universe to above $T_{\rm sph}$; see \Cref{sec:PT} for more details. 

For larger masses instead, kaon mixing constraints can exclude additional parameter space compared to $\mu \rightarrow e \gamma$.
We refer the reader to~\appref{app:flavorbounds} for more details, and here simply emphasize that for favorable charge assignments and $m_s \gtrsim 5 $ GeV, part of the viable parameter space of~\Cref{fig:Leptonic sector} can remain viable even when charging all fermions under $U(1)_{\rm FN}$. Independently of whether all SM fermions are charged under the $U(1)_{\rm FN}$, larger radial mode masses allow for smaller flavon VEVs. This implies less required supercooling, and therefore populating high-mass regions is anyways expected to be more natural than populating the low-mass counterpart, which becomes excluded in the case of an FN symmetry realized within all fermions.
\subsection{Testability}
As can be seen in \Cref{fig:Leptonic sector}, the viable parameter points live in a narrow band, bounded from below by flavor observables, and bounded from above by requiring the FNPT not to reheat the Universe above $T_{\rm sph}$.\footnote{There is some freedom in this bound since the reheating can be delayed, and a small period in which the washout is efficient can be countered by a larger initial asymmetry; see for details~\Cref{sec:PT}, specifically footnotes \,\ref{footnote:matDom}~and~\ref{footnote:reheating}. Still, it cannot be pushed up infinitely.}  However, as the branching ratio is proportional to $1/v_\Phi^4$, to probe one additional order of magnitude in the $m_\rho - v_\Phi$ plane, the constraint has to be improved by four orders of magnitude. Specifically, to fully test the parameter space via $\mu \rightarrow e \gamma$, the sensitivity will have to be improved to be $\sim 6 \times 10^{-17}$. The current leading constraints come from the MEGII experiment with a targeted final sensitivity of $6 \times 10^{-14}$ \cite{MEGII:2023fog}. 

As experimental precision improves, other lepton flavor violating observables will become relevant as well. The Mu3e experiment is designed to probe $\textrm{Br}(\mu \rightarrow 3e)$ down to a sensitivity of $10^{-16}$, improving previously existing bounds by four orders of magnitude \cite{Mu3e:2020gyw,Hesketh:2022wgw}. The BelleII experiment plans to obtain a sensitivity of $\textrm{Br}(\tau \rightarrow 3 \mu) = 3.3 \times 10^{-10}$ \cite{Belle-II:2018jsg}, competitive with $\mu\rightarrow 3e$ due to the larger tau-flavon couplings. Furthermore, in the presence of flavon-quark couplings, $\mu \rightarrow e$ conversion in nuclei can become important. The Mu2e experiment \cite{Mu2e:2014fns} and the COMET experiment \cite{COMET:2018auw} will look for $\mu \rightarrow e$ conversion in aluminum nuclei, improving the current limit by more than four orders of magnitude to $\sim 10^{-17}$. Both plan to start data taking in 2027 \cite{Driutti:2026pqo,Clouvel:2026hrt}.

Thus, as constraints on flavor violation improve, the region in which the model remains viable will be probed progressively. However, the experiments currently under construction will not yet probe the full parameter space, only barely surpassing the current $\mu \rightarrow e \gamma$ bound.
%
\section{Phase transition dynamics}\label{sec:PT}
For successful sequestering of charged-lepton chiral asymmetries under the prescription described above, without conflicting with flavor observables, a hierarchy between the flavon VEV $v_\Phi$ and the FNSB temperature $T_\mathrm{FN}$ is required.\footnote{This setup is reminiscent but opposite to Ref.~\cite{Baldes:2016gaf}, where the Yukawas start off being very large in the early Universe due to the flavon having a \textit{larger} VEV at high temperatures.} 
In other words, the FNPT must be supercooled.

To understand the scalar field dynamics, we consider the simple potential
\begin{equation}
    V(\Phi) =V_0 + V_T + V_\mathrm{CW} + V_\mathrm{CT} + V_\mathrm{Daisy} \, .
    \label{eq:fullpot}
\end{equation}
$V_0$ the tree-level potential, for which we assume a negligible portal coupling between the flavon and the Higgs boson. $V_T$ is the 1-loop finite-temperature potential, containing contributions from all fields with which the flavon is in thermal equilibrium. It is given by
\begin{equation}
    V_T = \frac{T^4}{2 \pi^2} \sum_i n_i J_{B,F} \left( \frac{m_i^2(\Phi)}{T^2}\right)\, ,
    \label{eq:VT}
\end{equation}
with the functions $m_i^2$ denoting the (squared) field-dependent masses, given in~\appref{app:FDmasses}, and the functions $J_{B/F}$ for bosons (B) and fermions (F) given by
\begin{equation}
    J_{B/F} (y) = \pm \int_0^\infty x^2 \log{\left(1 \mp \exp{\!(-\sqrt{x^2 + y})}\right)} dx\,,
    \label{equ:thermalFcts}
\end{equation}
where the lower signs correspond to fermions and the upper ones to bosons.

Due to the supercooled nature of the phase transition, the minimum in field space is much larger than the phase transition temperature, so the high temperature approximation does not accurately capture the relevant dynamics. Therefore, in the following, we include the full thermal functions, which can be well approximated by a sum of Bessel functions of the second kind as \cite{Anderson:1991zb}
\begin{equation}
    J_{B,F} (y) = \mp \sum_{l=1}^{5} \left(\pm 1\right)^l \frac{y}{l^2} K_2(\sqrt{y}l)\,.
\label{equ:BesselFctApprox}
\end{equation}
There is one intricacy in this FN set-up that requires an additional ingredient in the model. Since the field-dependent masses of the SM fermions are zero in the $h=0$ direction, before EWSB they do not contribute to $V_T$ in the flavon potential. In addition, the flavon self-interaction is bounded to be small due to the supercooling requirement, which can only be satisfied in flat potentials. Thus, to ensure that the flavon potential has a minimum at $\Phi = 0$ before $T_{\rm sph}$,
we consider the minimal extension of the field content by an additional Weyl fermion $\xi$ with $U(1)_{\rm FN}$ charge satisfying the condition $[\overline{\xi^c} \xi]_{\rm FN} = 1$, such that the interaction
\begin{equation}
    \mathcal{L} \supset - y_\xi \Phi \overline{\xi^c} \xi \,,
    \label{eq:lagrangianF}
\end{equation}
is added to the Lagrangian. With values of the coupling $y_\xi\sim\mathcal{O}(1)$, $\xi$ will give the dominant contribution to the thermal potential before EWSB.\footnote{Alternatively, one could attempt to add restoring effects by gauging the $U(1)_{\rm FN}$ symmetry.}

Compared to the FN fermions appearing in a UV completion, which have large masses of $\mathcal{O}(\Lambda_\mathrm{FN})$, the fermion $\xi$ will acquire a mass $m_\xi \sim y_\xi v_\Phi<\Lambda_{\rm FN}$. The one-loop Coleman-Weinberg (CW) potential in the Landau gauge and $\overline{\mathrm{MS}}$-scheme is given by
\begin{equation}
    V_\mathrm{CW} = \sum_i n_i \frac{m_i^4}{64 \pi^2} \left( \ln \left(\frac{m_i^2}{Q^2} \right)-C_i\right)\,,
    \label{eq:VCW}
\end{equation}
where the sum includes all particles in the thermal bath. The renormalisation scale $Q$ is chosen to be $Q=0.1\times \Lambda_\mathrm{FN} \simeq v_\Phi$ and $C_f=C_F=C_\Phi=3/2$. The degrees of freedom are $n_q = -12$, $n_l =-4$, $n_F=-2 $ and $n_\Phi = 2$. Since the CW potential can shift the minima, we include a counterterm potential
\begin{equation}
    V_\mathrm{CT} = \delta m \Phi^2 + \delta \lambda \Phi^4 \, ,
    \label{eq:VCT}
\end{equation}
with the coefficients $\delta m$ and $\delta\lambda$ determined from the conditions
\begin{equation}
    \frac{\partial V_\mathrm{CW}}{\partial \Phi} = - \frac{\partial V_\mathrm{CT}}{\partial\Phi} \quad , \quad \frac{\partial^2 V_\mathrm{CW}}{\partial \Phi^2} = - \frac{\partial^2 V_\mathrm{CT}}{\partial\Phi^2} \, .
    \label{eq:CTcond}
\end{equation}
This ensures that the values of the VEV and masses of the tree-level potential, evaluated at $\Phi = v_\Phi$ and $h = v$, are preserved after the CW contributions are accounted for. Lastly, we include the thermal corrections to the scalar potential from the resummation of Daisy diagrams as \cite{Carrington:1991hz,Arnold:1992rz}
\begin{equation}
    V_\mathrm{Daisy} = - \frac{T}{12 \pi} n_\Phi \left( M^3_\Phi - m_\Phi^3 \right) \, ,
    \label{eq:VDaisy}
\end{equation}
where $M_\Phi^2 = m_\Phi^2 + \Pi(T)$ is the thermal mass of the scalar field; see~\appref{app:FDmasses} for details.
\subsection{ Analytical analysis of scalar potential}
As the UV model includes heavy fields which are integrated out to obtain the fermion mass operators, we also consider higher-order operators in the scalar potential. For simplicity, we truncate the potential at dimension six
\begin{equation}
    V_0(\Phi) = -\mu^2 \Phi^2 - \lambda \Phi^4 + \frac{4c}{3 \Lambda_\mathrm{FN}^2} \Phi^6 \, ,
    \label{eq:V6}
\end{equation}
but for example dimension-eight operators, which would be further suppressed, could also be straightforwardly added. The signs are chosen for convenience.

Since the ALP plays a subdominant role in the present leptogenesis framework, we do not reduce the analysis by assuming a specific term. However, the explicit symmetry breaking (exSB) can also give a contribution to the mass term of the radial mode, imagine e.g. $V_\mathrm{exSB} \sim \mu_\mathrm{exSB}^2 \left( \Phi \Phi + \Phi^\dagger \Phi^\dagger\right)$. From now on, we assume that the contribution from the exSB is implicitly included into the quadratic term of the radial mode.

For the analytical estimate, we neglect $\mu^2$, which is required to be small in magnitude so that thermal corrections can lead to a positive curvature at the origin, enabling a thermal barrier to survive until $T_{\rm sph}$. The flavon VEV is given by
\begin{equation}
    v_\Phi = {\sqrt{\frac{\lambda}{c}} \,\Lambda_\mathrm{FN}}\, ,
\end{equation}
from which we can derive the condition $\lambda \sim 10^{-2}\, c$ in order to obtain $\epsilon \sim 0.1$. The radial mode after FNSB has a resulting mass of
\begin{equation}
    m_\rho = \sqrt{\frac{ {2 \lambda^2}}{c}} \Lambda_\mathrm{FN} \sim  \sqrt{c}\, \left(10^{-2}\,\Lambda_\mathrm{FN}\right)\, .
\end{equation}
The advantage of considering additional operators\footnote{In~\appref{app:nophi6} we showcase the situation for a scalar potential without higher dimensional operators. This has the disadvantage that $\mu^2$ is crucial for both the flavon mass and the temperature of the PT, and large masses come at the cost of breaking the FN symmetry before sphalerons have shut off, therefore washing out the asymmetry. Furthermore, many of the minima are only metastable.} is the ``freeing up" of $\mu^2$, which here only determines the phase transition temperature, while the flavon's mass and VEV depend primarily
on $\lambda$ and $c$. Thermal corrections will lead to an positive effective mass parameter $\mu^2_\mathrm{eff}>0$, thus introducing a thermal barrier into the scalar potential when $\left| \mu^2 \right|<y_\xi^2 T^2$. 
\subsection{Supercooling}
\begin{figure}
    \centering
    \includegraphics[width=0.99\linewidth]{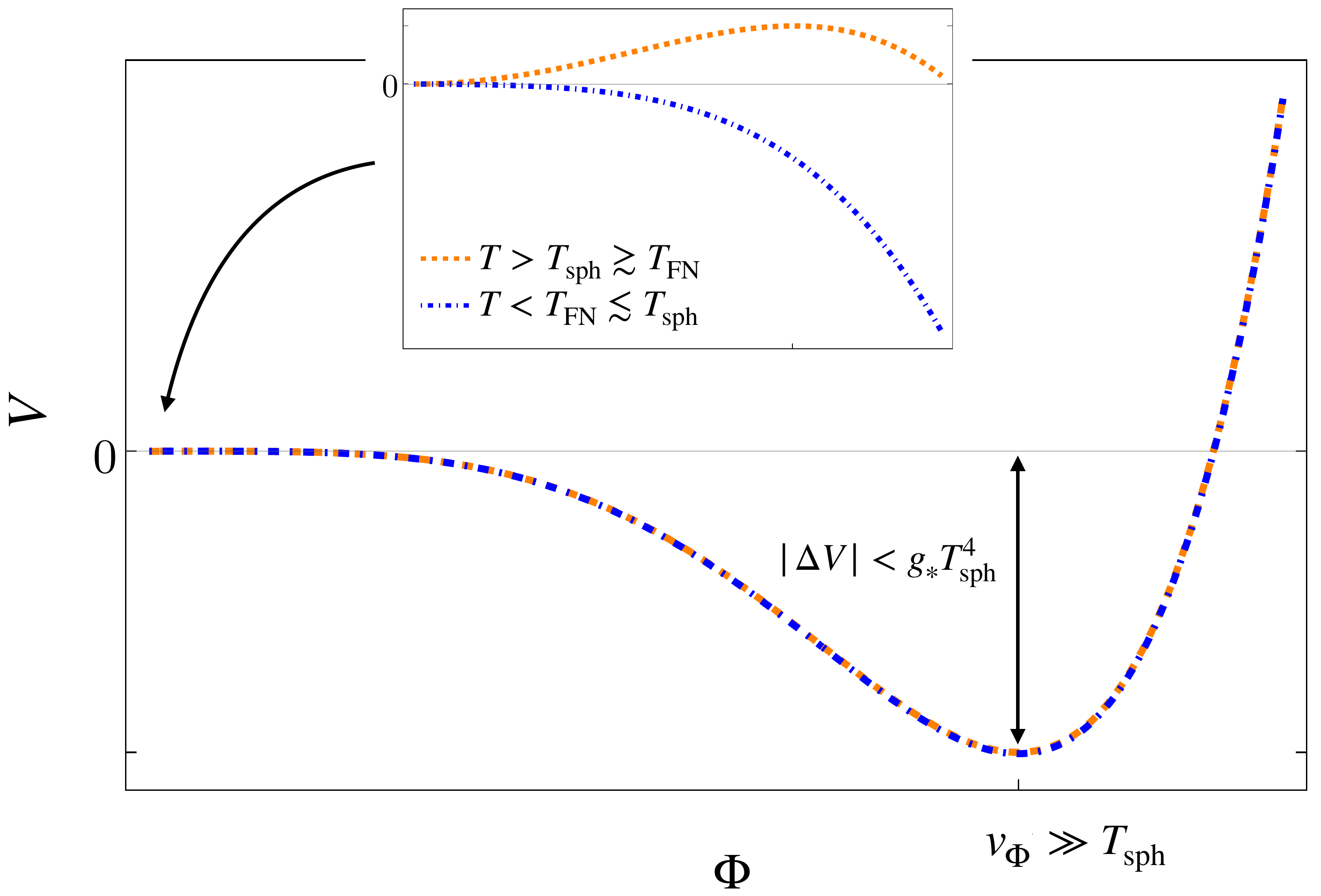}
    \caption{Illustration of the requirements the effective potential in \Cref{eq:fullpot} has to realize. Above the sphaleron temperature, a small barrier remains (orange), keeping the field trapped at the origin. The barrier disappears at some $T_{\rm FN} \lesssim T_{\rm sph}$ (blue), allowing the field to obtain a VEV. Although the VEV is much larger than EW scales, the potential has to be sufficiently shallow such that the Universe is not reheated to temperatures above sphaleron freeze-out once the phase transition completes.}
    \label{fig:supercoolingPotential}
\end{figure}

As we require $T_{\rm FN}\ll v_\Phi$, the universe might enter a vacuum-dominated era, called \textit{thermal inflation} (TI) \cite{Lyth:1995hj,Lyth:1995ka}. Since $\rho_\mathrm{rad} \sim a^{-4}$, for sufficiently long supercooling the vacuum energy $\Delta V \sim \mathrm{const.}$ will necessarily start dominating. Then, the duration of the TI can be expressed in terms of e-folds $N$ as
\begin{equation}
    T_{\rm FN}/T_\mathrm{TI} = e^{-N} \, ,
\end{equation}
where $T_\mathrm{TI}$ is the temperature at which the vacuum energy starts dominating over the radiation energy density.  
Once the PT completes at $T_{\rm FN}$, the stored energy will be released into the plasma\footnote{This is a conservative estimate. If the phase transition was first order, some of the energy would end up in gravitational waves. Additionally, if there was a period of matter domination of the scalar field, the reheating of the thermal bath would be delayed, allowing for a larger initial $\Delta V$.\label{footnote:matDom}} and the reheating temperature can be estimated as 
\begin{equation}
    \mathopen|\Delta V\mathopen| \, \sim \,\rho_\mathrm{rh} = \frac{\pi^2}{30} g_\ast T_\mathrm{rh}^4 \,\, \Rightarrow \,\,T_\mathrm{rh} \, \sim \,\left(\frac{30 \mathopen|\Delta V\mathopen|}{\pi^2g_\ast} \right)^{1/4} \, .
    \label{eq:DeltaV}
\end{equation}
To successfully sequester the asymmetry, sphaleron processes should remain inactive once the PT completes, and therefore the universe should not reheat above the EW temperature.\footnote{\label{footnote:reheating}Technically, if the universe reheated only to be slightly above the $T_{\rm EW}$, sphalerons would become active again only for a short amount of time, which might not be problematic. One can imagine that a larger than necessary asymmetry was produced and then diluted down to the observed value.} To avoid such a problematically high reheating temperature, $\Delta V$ has to be sufficiently small, which in turn decreases the tunneling probability. At the same time, if the potential was too shallow, the flavon would be problematically light with respect to lepton flavor violation observables, specifically rare lepton decays. As a consequence the second minimum necessarily becomes global already at $T\gg T_{\rm sph}$, and the critical temperature is much larger than the final FNPT temperature $T_{\rm FN}$.
\Cref{fig:supercoolingPotential} shows requirements in terms of the effective potential, highlighting the required magnitudes in VEV and $\Delta V$.

For an analytical estimate, we can combine \Cref{eq:DeltaV}, where the vacuum energy is given by $\Delta V =  V(v_\Phi)-V(0) \sim V_0(v_\Phi) = -\frac{1}{24} m_\rho^2 v_\Phi^2$, with the requirement $T_\mathrm{rh} < T_\mathrm{EW}$, to obtain an upper bound on the mass of the radial mode, namely
\begin{align}
    m_\rho^2 \lesssim\, &\frac{4\pi^2 g_\ast T_\mathrm{EW}^4}{5v_\Phi^2} \nonumber\\\sim \, &\left( 50 \, \text{GeV}\right)^2 \left( \frac{g_\ast}{100} \right) \left( \frac{10^4 \, \text{GeV}}{v_\Phi}\right)^2 \, .
    \label{eq:reheatingUpperBound}
\end{align}
Of course, the mere existence of a barrier does not guarantee delayed FNSB, so as a next step we will calculate tunneling probabilities, using the full thermal potential.

\subsection{Tunneling probability}
Since we require $U(1)_\mathrm{FN}$ to be broken at the very latest before BBN, where we already expect to have massive fermions, we have to ensure that tunneling or quantum fluctuations are efficient enough to bring the field to the global minimum, or that the barrier disappears before then. The thermal tunneling rate is given by \cite{Coleman:1977py,Linde:1980tt,Linde:1981zj}
\begin{equation}
    \Gamma = T^4 \left( \frac{S_3}{2 \pi T}\right)^{3/2} e^{-S_3/T}\,,
\end{equation}
and once this becomes comparable to the Hubble rate, bubbles start nucleating. The nucleation temperature is defined by the moment when one bubble per Hubble volume has formed, namely
\begin{equation}
    \int_{t_c}^{t_n} dt \frac{\Gamma(t)}{H(t)^3} = 1 \, ,
\end{equation}
where the integral spans from the critical time $t_c$, i.e. when the minima are equally favorable, to the nucleation time $t_n$. 

To calculate the action, we follow the tunneling potential formalism from Ref.~\cite{Espinosa:2018hue}, and in the relevant parameter space the action becomes small, i.e. nucleation becomes favorable, just before the barrier disappears completely. In that sense, our setup differs from what is often done in supercooled phase transitions; see e.g. Ref.\,\cite{Lewicki:2021xku}. The potential is not (quasi)conformal, but contains an explicit negative $\mu^2$ term which will always start dominating over thermal corrections, thus ensuring that the field cannot get stuck in the unbroken phase. This makes the competition between expanding bubbles and the expanding Universe due to the vacuum domination less crucial. If some bubbles form, there is very little time before the barrier disappears and the phases start mixing. Thus, we consider the temperature at which the barrier disappears to be the temperature at which the phase transition occurs, $T_{\rm FN}$, and treat it as a second-order phase transition. Since $T_{\rm FN}$ is much smaller than the critical temperature, the phase transition is supercooled. 

A more detailed analysis of the model, particularly keeping possible gravitational wave signals in mind, goes beyond the scope of this paper.\footnote{Ref.~\cite{Dutka:2025oqt} carried out a study including lattice simulations about supercooled phase transitions with a curvature flip.} In addition, since the here-presented model contains an asymmetry in the charged leptons, primordial magnetic fields may be sourced \cite{Joyce:1997uy}. Both of these questions are left for future work.
\subsubsection*{Numerical scan}
In order to present numerical results, we scan over the input parameters $\mu^2,\lambda,c,\Lambda_{\rm FN}, c_f,y_\xi$ in the ranges specified in~\Cref{tab: scan inputs} and filter for a phase transition occurring between $T_{\rm sph}$ and BBN. Here, we continue using the simplifying assumption of diagonal Yukawas, as for the leptogenesis analysis. Working with the full textures is essential only when determining the bounds from flavor violation. The values of $c_f$ are then obtained in terms of $y_f^{\rm SM}$, the SM Yukawas, such that the generated fermion masses match observations. 

\Cref{fig:Leptonic sector} shows the successful points in the $m_\rho$ - $v_\Phi$ plane together with the constraints from flavor violation. The numerically determined reheating temperature is in good agreement with the analytical estimate from requiring a viable reheating temperature per \Cref{eq:reheatingUpperBound}. The red (green) scatter points satisfy the condition $T_{\rm rh}<T_{\rm sph}$ ($T_{\rm rh}<T_{\rm sph}/2$), and the solid lines are obtained from the analytical estimate. 
\begin{table}[t]
\centering
\setlength{\tabcolsep}{3pt}
\begin{adjustbox}{max width=\columnwidth}
\begin{tabular}{@{}cc@{\hspace{0.95em}}cc@{}}
\hline
Parameter & Range & Parameter & Value \\
\hline
$\mu^2$ & $[10^{-8},10]\,\mathrm{GeV}^2$
  & $\epsilon$ & $0.1$ \\
$\Lambda_{\rm FN}$ & $[10^4,10^{7}]\,\mathrm{GeV}$
  & $\lambda$ & $-\mu^2/(\sqrt{2}\epsilon\Lambda_{\rm FN})^2
                    +2c\epsilon^2$ \\

  $c$ & $[10^{-15},1]$&$n_f$& \Cref{tab: FN eps=0.1}  \\
$y_\xi$ & $[10^{-8},1]$
  & $c_f$ & $y_f^{\rm SM}/\epsilon^{n_f}$ \\
\hline
\end{tabular}
\end{adjustbox}
\caption{Inputs for the flavon phase-transition scan. Parameters on the left are sampled uniformly in $\log_{10}$ of their values. The right block gives fixed and derived parameters.}
\label{tab: scan inputs}
\end{table}
\subsection{Dilution}
\begin{figure}
    \centering
    \includegraphics[width=0.99\linewidth]{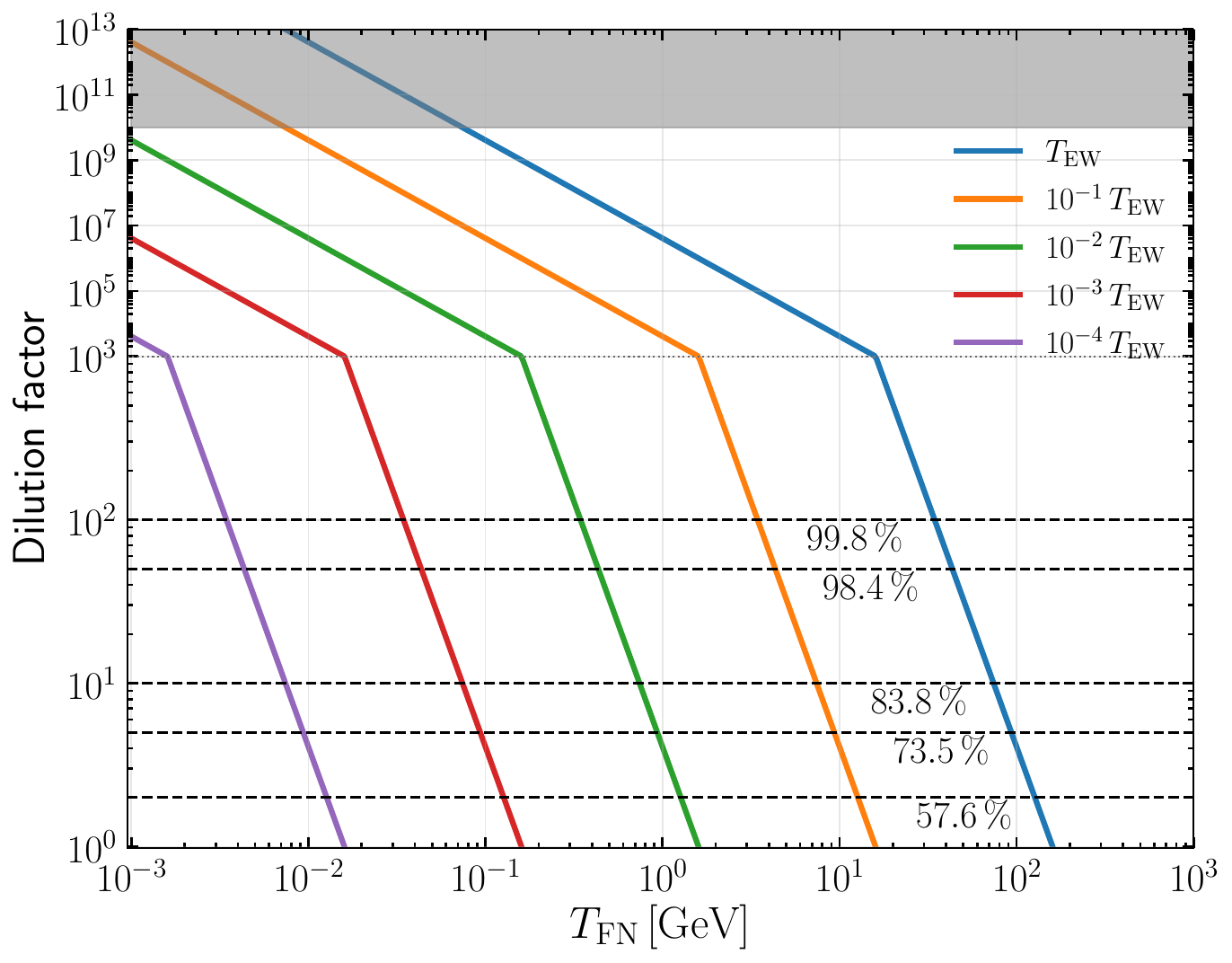}
    \caption{Dilution factor as a function of the temperature at which the phase transition completes, i.e. when thermal inflation ends. The colored lines correspond to different reheating temperatures, and a dilution $>10^{10}$ is phenomenologically excluded since even an initial $\mathcal{O}(1)$ asymmetry would not be able to match observations. The dashed lines indicate how many of the points in the numerical scan lie below a certain dilution.}
    \label{fig:dilutionFactor}
\end{figure}
When there is reheating of the thermal bath, the baryon yield will be diluted, independently of sphaleron processes. The dilution will be estimated in the following. 

The baryon asymmetry yield can be expressed as
\begin{equation}
    Y_B = \frac{n_B - n_{\Bar{B}}}{s}\,,
\end{equation}
where $s \sim g_\ast T^3$ is the entropy density. By reheating the bath from $T_{\rm FN}$ to $T_\textrm{rh}$, the yield will be diluted according to
\begin{equation}
    \frac{Y^\text{in}_{B}}{Y^\text{fin}_{B}} = \frac{s_\mathrm{fin}}{s_\mathrm{in}}\sim \left( \frac{T_\mathrm{rh}}{T_\textrm{FN}}\right)^3 \, .
    \label{eq:dilution}
\end{equation}
As $\Delta V \sim \rho_\mathrm{rh}$, see~\Cref{eq:DeltaV}, the reheating temperature can be determined directly from the scalar potential.~\Cref{fig:dilutionFactor} shows the dilution factor (\Cref{eq:dilution}) for different reheating temperatures, as a function of $T_{\rm FN}$, after which the PT completes and thermal inflation ends. We note it is possible that the value of the initially injected asymmetry may have been higher than expected, and diluted by the additional entropy to its nowadays value. We set the maximal initially generated asymmetry yield to order one, in which case the absolute upper bound on the dilution factor is of $\mathcal{O}(10^{10})$, to still be able to match the observed BAU~\cite{Planck:2015fie,Planck:2015fie,Cooke:2017cwo,Yeh:2022heq}
\begin{equation}
   \eta_B =\frac{n_B - n_{\bar{B}}}{n_\gamma} \sim 6\times10^{-10} \,,
\end{equation}
which can be equivalently written in terms of the baryon asymmetry yield $Y_B=(n_B - n_{\bar{B}})/s\sim 9\times 10^{-11}$. 

In the numerical scan, the reheating for all but $0.2\%$ of the points led to a dilution factor $\leq 10^2$.\footnote{This can be understood intuitively; it is already somewhat tricky to obtain enough supercooling to delay the PT until after sphalerons have frozen out, and large dilutions can only occur for $T_{\rm FN} \ll T_\mathrm{rh}\, \lesssim \, T_{\rm sph}$, i.e. even larger supercooling.} In this case, an initial asymmetry is required of at most two orders of magnitude larger than expected without additional entropy injection. For consistency reasons, we require the FNSB to happen before BBN, which places a lower bound on $T_{\rm FN}$. In general, for large reheating close to the EWPT, the FNPT has to complete before $T_{\rm FN}\gtrsim 0.1$ GeV, while for smaller reheating temperatures there is no problematically high dilution. 
\section{Conclusion and Outlook}\label{sec:conclusion}
We have shown that Dirac leptogenesis, which does not rely on lepton number violation to source an initial lepton asymmetry, can proceed via all charged leptons, and not only Dirac neutrinos with tiny Yukawa couplings as originally proposed in Ref.~\cite{Dick:1999je}. 

In fact, the desired dynamics can arise in frameworks explaining the fermion mass hierarchies from a UV origin, such as a Froggatt-Nielsen mechanism. There, the low-energy Yukawa couplings, which normally lead to a washout of the asymmetry for the charged leptons before sphaleron freeze-out, are only induced once the scalar field associated with the $U(1)_\textrm{FN}$ symmetry breaking obtains a VEV. Before, the left-right equilibration proceeds via higher-dimensional operators which \textit{exit} equilibrium as the Universe cools down since they become thermally suppressed with $(T/\Lambda_\mathrm{FN})^{2 n_f}$ as the temperature drops below the eventual flavon VEV. Thus, the supercooling of the FN phase transition until sphaleron freeze-out opens a temperature range in which a chiral lepton asymmetry can be sourced, successfully remain sequestered by the leptons, and converted to a baryon asymmetry by EW sphalerons. 

To assess the viability of this framework, we estimated the thermally averaged chirality-equilibration rates and identified the allowed asymmetry-production temperatures. Explicitly, with the presented benchmark, successful sequestering requires $\Lambda_\textrm{FN}>10^5\,\text{GeV}$ for the tau lepton, $\Lambda_\textrm{FN}>10^4\,\text{GeV}$ for the muon and $\Lambda_\textrm{FN}>10^3\,\text{GeV}$ for the electron for the mechanism to work. However, there are additional, more stringent lower bounds on $\Lambda_{\rm FN}$, since FN constructions intrinsically lead to flavor violation. Since the described setup requires that the FNSB has to be delayed until sphalerons have shut off, this leads to a hierarchy $v_\Phi \gg T_{\rm FN}$. 

We studied the dominant lepton flavor violation process $\mu \rightarrow e \gamma$ as well as the finite-temperature flavon potential while also accounting for transition completion, reheating, and the resulting dilution of the primordial asymmetry. Combining these requirements, a light flavon is expected with a mass in the range $1.7 - 50\,\textrm{GeV}$, and where the allowed VEV is constrained from both above and below, leading to precise predictions which could be progressively testable via future lepton flavor violation experiments. 
Finally, a more detailed treatment of the phase transition dynamics could furthermore establish whether bubble formation occurs, results in observable gravitational-wave signatures and whether primordial magnetic fields are sourced.
%
\section{Acknowledgments}
We are grateful to Admir Greljo, Yann Gouttenoire, Joerg Jaeckel, Mark Hindmarsh, Manfred Lindner, Maura E. Ramirez-Quezada, Aleks Smolkovic, Carlo Tasillo and Jorinde van de Vis for useful discussions. J. P. G. acknowledges funding from the International Max Planck Research School for Precision Tests of Fundamental Symmetries (IMPRS-PTFS). MH and FG would like to express special thanks to the Mainz Institute for Theoretical Physics (MITP)
of the Cluster of Excellence PRISMA+ (Project ID 390831469), for its hospitality and support during the completion of this work.
\appendix
\section{Left-right equilibration}\label{sec:App A}
\subsection{Thermally averaged rates: fully unbroken phase}\label{sec: th rates unb phase}
We aim to estimate the contributions of the $2\to n_f+1$ processes in the top row of~\Cref{fig:FN diagrams} to the thermally averaged rate of left-right equilibration processes
\begin{equation}\label{eq: Gamma LR Ap}
    \langle \Gamma_{\rm LR}^f\rangle(T)=\frac{\gamma_{\rm LR}^f(T)}{n_{f}^{\rm eq}(T)}\,,
\end{equation}
where $n_{f}^{\rm eq}(T)$ is the equilibrium number density of fermion $f$ at temperature $T$ and
\begin{equation}\label{eq: gamma LR}
    \gamma_{\rm LR}^f(T)=\frac{g_1 g_2 T}{32\pi^4}
\int_{s_{\rm min}}^\infty ds\,\sqrt{s}\,\hat\sigma^f_{\mathrm{LR}}(s)\,K_1\!\left(\frac{\sqrt{s}}{T}\right)\,,
\end{equation}
assuming Maxwell-Boltzmann (MB) statistics. $K_1$ denotes the modified Bessel function of the first kind, $g_i$ (i=1, 2) denotes the internal degrees of freedom of the initial-state particles, and 
\begin{equation}
    s_{\rm min}=\max\left\{(m_1+m_2)^2,\,\left(\sum_{i=3}^{n_f+3}m_i\right)^2\right\}\,,
\end{equation}
with $m_i$ denoting masses of initial state particles for $i=1,2$ and of final state particles for $i\geq 3$. The reduced cross-section is defined as
\begin{equation}\label{eq: hat sigma LR}
    \hat \sigma^f_{\mathrm{LR}}(s)=\frac{\lambda(s,\,m_1^2,\,m_2^2)}{s}\,\sigma^f_{\mathrm{LR}}(s)\,,
\end{equation}
with 
\begin{align}
    \lambda(s,\,m_1^2,\,m_2^2)=\,&s^2+m_1^4+m_2^4-2\,s\,m_1^2\nonumber\\ &-2\,s\,m_2^2-2\,m_1^2\,m_2^2\,,
\end{align}
and the cross section $\sigma$ is obtained from the phase space integral of the squared matrix element associated with the left-right equilibration processes, namely
\begin{equation}
    \sigma^f_{\mathrm{LR}}(s)=\frac{1}{2\,s}\int d\Phi\,|\mathcal{M}^f_{\mathrm{LR}}|^2\,.
\end{equation}

%
\subsubsection{Left-right annihilation}\label{sec: app LR annihilation unb}
For the annihilation process $f_L(p_1)+\bar f_R(p_2)\to H(k_0)+\sum_{i=1}^{n_f}\Phi(k_i)$ we have
\begin{equation}\label{eq: annihilation amplitude squared}
    |\mathcal{M}^f_{\mathrm{LR}}|_s^2=\frac{(n_f!)^2|c_f|^2}{\Lambda_{\mathrm{FN}}^{2n_f}}\,(s-2\,m_f^2)\,,
\end{equation}
which can be factored out of the phase space integral, namely
\begin{equation}\label{eq: sigma LR}
    \sigma_{\mathrm{LR},\,s}^f(s)=\frac{1}{2\,s}\,\frac{1}{n_f!}\,|\mathcal{M}^f_{\mathrm{LR}}|_s^2\,\Phi_{n_f+1}\,,
\end{equation}
where we have included the symmetry factor due to identical final-state particles, and defined
\begin{align}\label{eq: phase space}
    \Phi_{n_f+1}=\int(2\pi)^4 \delta^{(4)}\!\left(p_{in} - k_f\right)\prod_{i=0}^{n_f}\frac{d^3 k_i}{(2\pi)^3 2E_i}\,,
\end{align}
with $p_{in}=p_1+p_2$, the total initial momentum, and $k_f=\sum_{i=0}^{n_f} k_i$, the total final momentum.

In the limit $s\gg\{m_i\}$ the integral of~\Cref{eq: phase space} can be solved analytically. We start by performing the factorization
\begin{align}
    d\Phi_{n_f+1}(s;k_0,...,k_{n_f})=&\frac{dq^2}{2\pi}d\Phi_2(s;q,k_0)\nonumber\\
    &\times d\Phi_{n_f}(q;k_1,...,k_{n_f})\,,
\end{align}
with $q^\mu=\sum_{i=1}^{n_f}k_i^\mu$. Integrating over $\{k_i\}$ one obtains
\begin{equation}\label{eq: Phi nf+1}
    \Phi_{n_f+1}(s)=\frac{1}{2\pi}\int dq^2\Phi_2(s;q^2,0)\,\Phi_{n_f}(q^2)\,,
\end{equation}
where the two-body phase space reads
\begin{equation}\label{eq: Phi 2}
    \Phi_2(s;q^2,0)=\frac{\sqrt{\lambda(s;q^2,0)}}{8\pi s}=\frac{s-q^2}{8\pi s}\,.
\end{equation}
Next, we take the inductive ansatz
\begin{equation}\label{eq: Phi nf ansatz}
    \Phi_{n_f}(q^2)=\frac{(q^2)^{n_f-2}}{2(4\pi)^{2n_f-3}\Gamma(n_f)\Gamma(n_f-1)}\,,
\end{equation}
with $\Gamma(n)=(n-1)!$, and for which $n_f=2$ gives the correct value $\Phi_2(s)=1/8\pi$. Substituting~\Cref{eq: Phi 2,eq: Phi nf ansatz} into~\Cref{eq: Phi nf+1} then gives
\begin{align}
    \Phi_{n_f+1}(s)=&\frac{1}{2\pi}\frac{1}{8\pi s}\,I_1(s)\nonumber\\&\times\frac{1}{2(4\pi)^{2n_f-3}\Gamma(n_f)\Gamma(n_f-1)}\,,     
\end{align}
with
\begin{equation}
    I_1(s)=\int_0^s dq^2\,(s-q^2)(q^2)^{n_f-2}=\frac{s^{n_f}}{(n_f-1)\,n_f}\,.
\end{equation}
Thus, we obtain
\begin{equation}
    \Phi_{n_f+1}(s)=\frac{s^{(n_f+1)-2}}{2(4\pi)^{2(n_f+1)-3}\Gamma(n_f+1)\Gamma((n_f+1)-1)}\,,\nonumber
\end{equation}
which confirms the inductive hypothesis. Substituting this result in~\Cref{eq: sigma LR}, then in~\Cref{eq: hat sigma LR} and finally in~\Cref{eq: gamma LR} gives
\begin{equation}\label{eq: gamma LR I2}
    \gamma_{\mathrm{LR},\mathrm{s}}^f(T)=\frac{|c_f|^2}{8\,(4\pi)^{2n_f+1}\,\Gamma(n_f)}\frac{T}{\Lambda_{\mathrm{FN}}^{2n_f}}\,I_2(T)\,,
\end{equation}
with
\begin{equation}
    I_2(T)=\int_0^\infty ds\,s^{n_f+1/2}K_1\left(\frac{\sqrt{s}}{T}\right)\,.
\end{equation}
With the change of variables $x=\sqrt{s}/T$ the integral can be rewritten as
\begin{equation}
    I_2(T)=2\,T^{2n_f+3}\,\int_0^\infty dx\, x^{2n_f+2}\,K_1(x)\,,
\end{equation}
and using the identity
\begin{align}
    \int_0^\infty dx\,x^\mu K_\nu(x)=2^{\mu-1}&\Gamma\!\left(\frac{1+\mu+\nu}{2}\right)\nonumber\\
    &\times\Gamma\!\left(\frac{1+\mu-\nu}{2}\right)\,
\end{align}
for $\mu=2n_f+2$ and $\nu=1$, we obtain
\begin{equation}
\int_0^\infty dx\,x^{2n+2}K_1(x)=2^{2n_f+1}\Gamma(n_f+2)\Gamma(n_f+1).
\end{equation}
Substituting these results in~\Cref{eq: gamma LR I2} and then on~\Cref{eq: Gamma LR Ap}, and using the MB equilibrium number density
\begin{equation}
    n_f^{\mathrm{eq}}=\frac{T^3}{\pi^2}\,,
\end{equation}
gives the final result for the thermally-averaged tree-level left-right annihilation rate for fermion $f$ in the limit $s\gg\{m_i\}$, namely
\begin{equation}\label{eq: Gamma LR s}
    \langle\Gamma_{\mathrm{LR},\,s}^f\rangle(T)=\kappa_f^{d>4}\,|c_f|^2\,\frac{T^{2n_f+1}}{\Lambda_{\mathrm{FN}}^{2n_f}}\,,
\end{equation}
with
\begin{equation}
    \kappa_f^{d>4}=\frac{n_f\,(n_f+1)!}{4\,(2\pi)^{2n_f+1}}\,.
\end{equation}
\subsubsection{Left-right scattering}
For the scattering process $f_L(p_1)+H(p_2)\to f_R(k_0)+\sum_{i=1}^{n_f}\Phi(k_i)$ we have
\begin{equation}
    |\mathcal{M}^f_{\mathrm{LR}}|_t^2=2\frac{(n_f!)^2|c_f|^2}{\Lambda_{\mathrm{FN}}^{2n_f}}(p_1\cdot k_0)\,,
\end{equation}
which gives the cross-section
\begin{equation}
    \sigma_{\mathrm{LR},\,t}^f(s)=\frac{n_f!\,|c_f|^2}{s}\frac{p_\mu}{\Lambda_{\mathrm{FN}}^{2n_f}}\,\int d\Phi_{n_f+1}k_0^\mu\,.
\end{equation}
In order to evaluate the phase-space vector integral, we note that, due to Lorentz covariance, it must be proportional to $P^\mu\equiv p_1^\mu+p_2^\mu$, namely
\begin{equation}
    \int d\Phi_{n_f+1}k_0^\mu=A(s)\,P^\mu\,.
\end{equation}
Contracting both sides with $P_\mu$ gives
\begin{equation}
    \int d\Phi_{n_f+1}\,(P\cdot k_0)=A(s)\,P^2=A(s)\,s\,,
\end{equation}
and in the limit $s\gg\{m_i\}$, the final-state particles can all be treated on equal footing, allowing us to write
\begin{align}
    \int d\Phi_{n_f+1}\,(P\cdot k_i)=&\frac{1}{n_f+1}\int d\Phi_{n_f+1}\,\left(P\cdot \sum_{i=0}^{n_f}k_i\right)\nonumber\\
    =&\frac{s}{n_f+1}\Phi_{n_f+1}(s)\,,
\end{align}
where in the last equality we have used $P\cdot\sum_{i=0}^{n_f}k_i=P^2=s$. Thus, the unknown function $A(s)$ reads
\begin{equation}
    A(s)=\frac{1}{n_f+1}\Phi_{n_f+1}(s)\,,
\end{equation}
and the cross-section is given by
\begin{equation}
    \sigma_{\mathrm{LR},\,t}^f(s)=\frac{n_f!\,|c_f|^2}{2\,(n_f+1)}\frac{\Phi_{n_f+1}(s)}{\Lambda_{\mathrm{FN}}^{2n_f}}\,,
\end{equation}
where we have used $p_1\cdot P\simeq s/2$.

Finally, the thermally-averaged tree-level left-right scattering rate for fermion $f$ in the limit $s\gg\{m_i\}$ can be written in terms of its annihilation counterpart (\Cref{eq: Gamma LR s}) as
\begin{equation}
    \langle\Gamma_{\mathrm{LR},\,t}^f\rangle(T)=\frac{1}{n_f+1}\,\langle\Gamma_{\mathrm{LR},\,s}^f\rangle(T)\,.
\end{equation}
\subsubsection{Thermally averaged rates: $U(1)_{\mathrm{FN}}$ in broken phase}\label{sec: app LR annihilation brok}
Below $T_{\mathrm{FN}}$, the flavon obtains a non-vanishing VEV, and the effective square vertices of~\Cref{fig:FN diagrams} must be included in the computation of the left-right equilibration rate. These can mediate s- and t-channel annihilation processes such as $f_L+\bar f_R\to H$, $f_L+\bar f_R\to t+\bar t$, and $f_L+\bar f_R\to H+V_\mu$, where $V_\mu$ stands for a vector boson, as well as their scattering counterparts, i.e., $f_L+t\to f_R+\bar t$, and $f_L+H\to f_R+V_\mu$. 

Consider the process $f_L+\bar f_R\to H$. By detailed balance, we have
\begin{align}
    \gamma(f_L+\bar f_R\to H)=&\gamma(H\to f_L+\bar f_R)\\=&n_H^{\mathrm{eq}}\,\Gamma(H\to f_L+f_R)\frac{K_1(m_H/T)}{K_2(m_H/T)}\,,\nonumber
\end{align}
where
\begin{equation}
    n_H^{\mathrm{eq}}=g_H\frac{m_H^2\,T}{2\,\pi^2}\,K_2\left(\frac{m_H}{T}\right)\,,
\end{equation}
and
\begin{equation}
    \Gamma(H\to f_L+f_R)=\frac{|y_{\mathrm{eff}}^f|^2}{16\,\pi}\,m_H\,.
\end{equation}
Thus, we have
\begin{equation}
    \langle\Gamma_{\mathrm{LR}}\rangle(T)=\frac{g_H}{32\,\pi}|y_{\mathrm{eff}}^f|^2\frac{m_H^3}{T^2}\,K_1\left(\frac{m_H}{T}\right)\,.
\end{equation}
For $T\gg m_H$ we can approximate $K_1(m_H/T)\approx T/m_H$, and taking $m_H(T)\sim c_H\,T$ we obtain
\begin{equation}
    \langle\Gamma_{\mathrm{LR}}\rangle(T)\simeq \kappa_f^{d=4}\,|y_{\mathrm{eff}}^f|^2\,T\,,
\end{equation}
with
\begin{equation}
    \kappa_f^{d=4}=\frac{c_H^2}{32\,\pi}\,.
\end{equation}
Other processes modify $\kappa_f^{d=4}$ by factors of $\mathcal{O}(1)$. In particular, for the annihilation process $f_L+\bar f_R\to t+\bar t$ and $f_L+\bar f_R\to H+V_\mu$, as well as their scattering counterparts, one obtains $\langle\Gamma_{\mathrm{LR}}\rangle\sim y_t^2\,|y_{\mathrm{eff}}^f|^2\,T$ and $\langle\Gamma_{\mathrm{LR}}\rangle\sim g^2\,|y_{\mathrm{eff}}^f|^2\,T$, respectively, with $y_t$ the top Yukawa and $g$ a gauge coupling. 
\subsection{Effects of thermal masses}\label{sec: th mass effects}
The results in~\Cref{sec: th rates unb phase} are obtained assuming massless states, which allowed us to write analytical expressions for the thermally averaged rates of the left-tight equilibration processes. Importantly, even in the fully unbroken phase where $\langle \Phi\rangle=\langle H\rangle=0$, particles still acquire thermal masses due to interactions with the plasma. More precisely, the mass of particle $a$ in the thermal bath is given by
\begin{equation}
    m_a^2(T)=m_{a,0}^2+\Pi_a(T)\,,
\end{equation}
where $m_{a,0}$ corresponds to the bare mass and $\Pi_a(T)$ to the finite temperature self-energy contribution. For the high temperatures that we are interested in, the bare masses can be safely neglected, and the masses of weakly coupled particles are well described by
\begin{equation}
    m_a(T)\sim c_i\,T\,,
\end{equation}
with $c_i$ a function of the relevant gauge, Yukawa, or scalar couplings. 

An important consequence of non-vanishing thermal masses is their effect on the kinematics of the equilibration processes. For instance, the process 
\begin{equation}\label{eq: ann process}
    f_L+\bar f_R\to H+n_f\,\Phi\,,
\end{equation}
requires
\begin{equation}
    \sqrt{s}>m_H(T)+n_f\,\Phi\,,
\end{equation}
which shows that thermal masses reduce the phase space accessibility, suppressing the corresponding rate. From this, we can already conclude that the inclusion of thermal masses, if relevant, will have the effect of allowing for a wider range of temperatures for which the asymmetry production could take place. In this sense, the computations made above are a conservative estimate. 

We define the ratio
\begin{equation}\label{eq: F}
    F(T)\equiv \frac{\Gamma_{\rm LR}^{m\neq0}(T)}{\Gamma_{\rm LR}^{m=0}(T)}\,,
\end{equation}
to quantify the difference between the rates in the massless limit and with the inclusion of thermal masses.

For the annihilation process in~\Cref{eq: ann process}, we can factor out the squared amplitude from the phase space integral, which in the presence of non-vanishing masses cannot be evaluated analytically. Thus, we evaluate the MB reaction density by Monte Carlo integration over the collision energy and final-state phase space. At each sampled energy, we generate a massless final state with RAMBO~\cite{Kleiss:1985gy} and rescale its momenta to accommodate the thermal masses while conserving energy and momentum. We use the same samples for the massive and massless calculations, weighting the massive contribution to account for the modified phase space, incoming kinematics, and squared amplitude. We then include the change in the equilibrium fermion density to obtain the rate ratio $F$ in Eq~\Cref{eq: F}.

\begin{figure}
    \centering
    \includegraphics[width=0.95\linewidth]{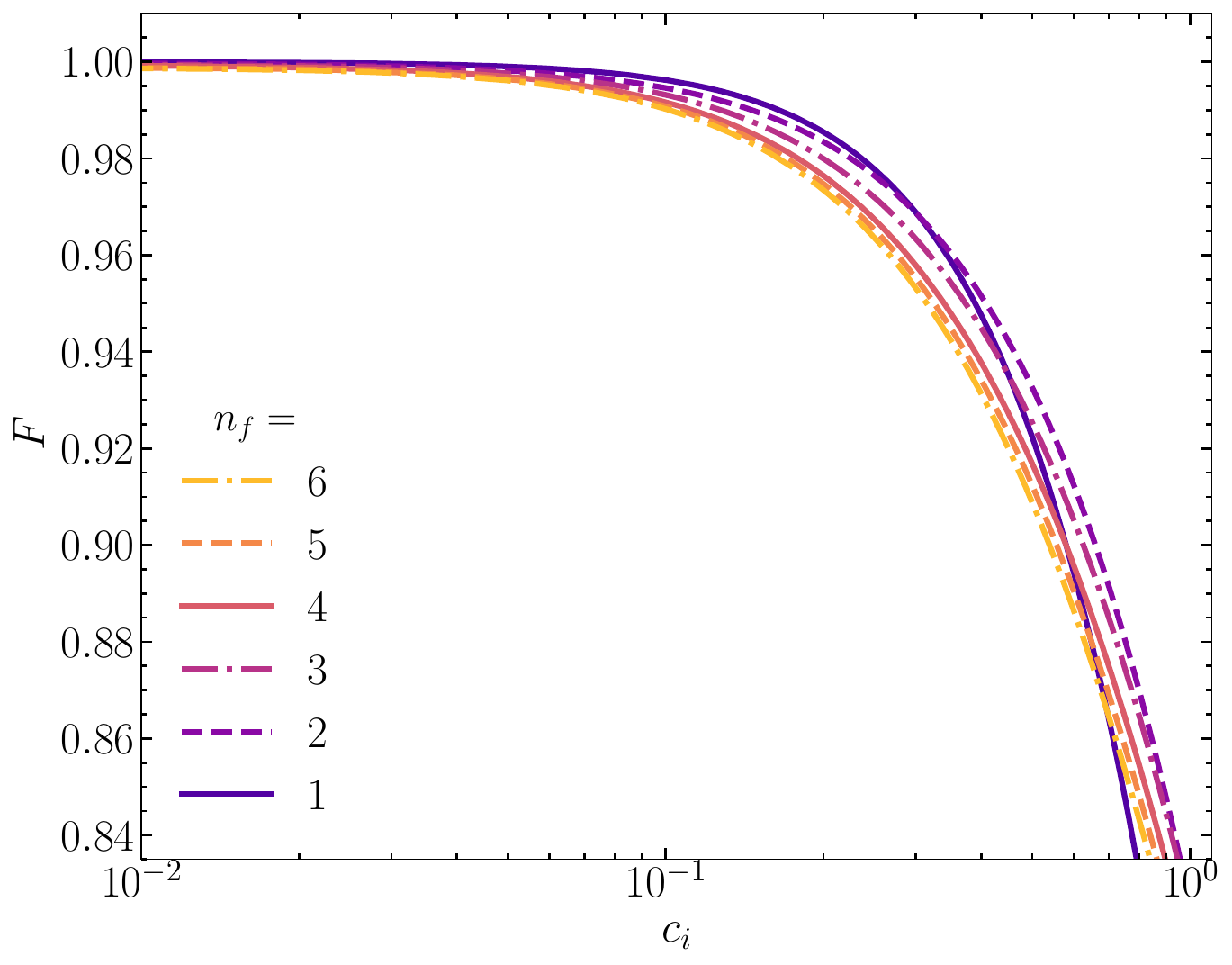}
    \caption{Fraction of thermally averaged rates with and without thermal masses included, for the processes $f_L+\bar f_R\to H+n_f\Phi$, as a function of different values of the thermal coefficients of the fermions and the Higgs.}
    \label{fig:F_ci}
\end{figure}

The resulting values of $F(T)$ are shown in~\Cref{fig:F_ci} for $c_i\in (0.1,1)$ for fermions and the Higgs boson, where gauge interactions are expected to give sizable contributions to the thermal masses, and $c_\Phi\sim 5\cdot 10^{-2}$ for the flavon, which is conservative since we assume the Higgs portal to be negligibly small when describing the FNPT. These results demonstrate that the effects of thermal masses to our observable of interest are at most of $\mathcal{O}(1)$. 
\subsection{Beyond Maxwell-Boltzmann}\label{sec: beyond MB}
We estimate the effect of quantum statistics assuming massless states and based on the s-channel processes $f_L+\bar f_R\to H+n_f\,\Phi$. For this, we define the ratio $R$ between the estimated rate using quantum statistics and massless states, $\gamma_{QS}^{m=0}$, and the rate in the massless MB approximation, $\gamma_{MB}^{m=0}$, namely
\begin{equation}
    R=\frac{\gamma_{QS}^{m=0}}{\gamma_{MB}^{m=0}}\,.
\end{equation}
\begin{figure}
    \centering
    \includegraphics[width=0.95\linewidth]{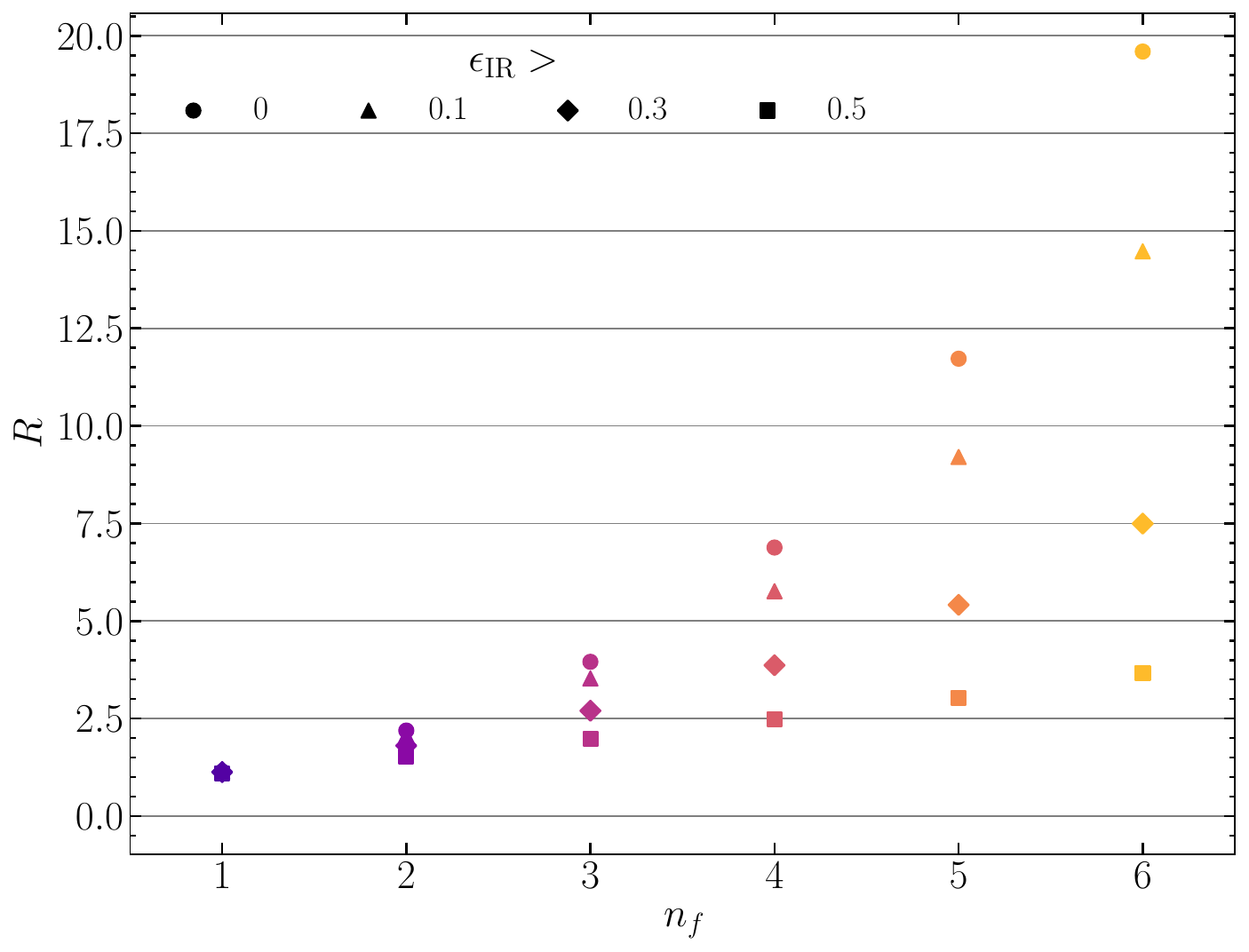}
    \caption{Effects of quantum statistics on the results obtained in the MB approximation with massless states, for different numbers of flavon in the final state of the process $f_L+\bar f_R\to H+n_f\, \Phi$. Different values of the infrared cutoff $\epsilon_{\rm IR}$, resembling final temperature effects, are included.}
    \label{fig:R_nf}
\end{figure}

At fixed $x=\sqrt{s}/T$, we use RAMBO~\cite{Kleiss:1985gy} to sample the massless $(n_f+1)$-body phase space uniformly. For each event $j$, we define $w_j(x)$ as its quantum-statistical weight relative to the MB weight. Writing $E_{ij}$ for the outgoing Higgs and flavon energies in the collision centre-of-mass frame, we have
\begin{align}
    w_j(x)=\frac{1}{(1+e^{-x/2})^2}\prod_{i=0}^{n_f}\frac{1}{1-e^{-E_{ij}/T}}\times\nonumber\\
    \mathbf{1}\left[\min_i(E_{ij}/T)\geq\epsilon_{\rm IR}\right]\,.
\end{align}
The first factor is the ratio of the incoming Fermi-Dirac and MB occupation products, and the product accounts for Bose enhancement of the outgoing bosons. The infrared cutoff sets the event weight to zero if any outgoing boson has $E_{ij}<\epsilon_{\rm IR}T$. Varying $\epsilon_{\rm IR}$ probes sensitivity to soft bosons within massless kinematics.

Since the squared amplitude in~\Cref{eq: annihilation amplitude squared} depends only on $s$, we estimate the phase-space average by the arithmetic mean $\overline{w}(x)=M^{-1}\sum_{j=1}^{M}w_j(x)$ over all $M$ generated events, including those with zero weight.  We approximate the quantum correction for all pair velocities by its value for a pair at rest in the plasma.\footnote{A full thermal average would evaluate each occupation factor at the plasma-frame energy $p_i\cdot u$, where $u$ is the plasma four-velocity, before integrating over the motion of the pair.} 

The results in~\Cref{fig:R_nf} show that, without an infrared cut, quantum statistics enhances the massless MB annihilation rate by an order of magnitude, more specifically factors of about $7$ to $20$, for $n_f>3$. The enhancement decreases as $\epsilon_{\rm IR}$ increases, indicating that soft final-state bosons contribute substantially. This means that the realistic ratio $R$ is expected to be smaller than the value computed with $\epsilon_{\rm IR}=0$, since e.g. thermal masses are expected to weaken Bose enhancement by raising the energies of such low-momentum modes.
\section{Phase transition dynamics}
\subsection{Field-dependent masses}\label{app:FDmasses}
For the potential in \Cref{eq:V6} the flavon field dependent mass is given by 
\begin{equation}
    m_\Phi^2 = 40 \frac{c}{\Lambda_\mathrm{FN}^2} \Phi^4-12 \lambda \Phi^2 -2 \mu^2\, .
\end{equation}
The thermal mass corrections in the Daisy potential in \Cref{eq:VDaisy} are
\begin{align}
    \Pi (T) = \frac{T^2}{12} \bigg( &-12 \lambda + 40 c \frac{\Phi^2}{\Lambda_\mathrm{FN}^2} \nonumber \\
    &+ 
    \frac{y_\xi^2}{2}  + \sum_i \frac{y_i^2}{2}   \frac{h^2\Phi^{2n_i -2}}{\Lambda_\mathrm{FN}^{2n_i}}\bigg)\, .
\end{align}
The SM fermions have field-dependent masses including both Higgs and flavon
\begin{equation}
    m_f^2 = \frac{1}{2} y_f^2 h^2 \frac{\Phi^{2n_f}}{\Lambda_\mathrm{FN}^{2n_f}} \, ,
\end{equation}
which, before EWSB, while $h = 0$ do not contribute to the flavon thermal potential, which motivated the inclusion of some additional fermions charged under $U(1)_\mathrm{FN}$ that couple to $\Phi$ directly, see \Cref{eq:lagrangianF} and accompanying description, with field-dependent mass 
\begin{equation}
    m_\xi^2 = \frac{1}{2} y_\xi^2 \Phi^2 \, .
\end{equation}

\subsection{Without $\Phi^6$ term}
\label{app:nophi6}
The scalar potential
\begin{equation}
     V_0 = -\mu^2 \Phi^2 + \lambda \Phi^4
     \label{eq:V04D}
\end{equation}
can be expressed in terms of vev $v_\Phi$ and radial mode mass after SB $m_0^2$ as
\begin{equation}
    V(\Phi) = -\frac{m_0^2}{2} \Phi^2 + \frac{m_0^2}{2 v_\Phi^2} \Phi^4\, .
    \label{eq:V04Dreexpressed}
\end{equation}
From this, $\Delta V = V(0)-V(v_\Phi) = - \frac{m_0^2}{8} v_\Phi^2$ follows. The reheating temperature can be estimated as
\begin{equation}
    \mathopen|\Delta V\mathopen| = \rho_\mathrm{rh} = \frac{\pi^2}{30} g_\ast T_\mathrm{rh}^4 \, \, \Rightarrow \, \, T_\mathrm{rh} = \left(\frac{30 \mathopen|\Delta V\mathopen|}{\pi^2g_\ast} \right)^{1/4} \, .
\end{equation}
and therefore
\begin{align}
    m_0^2 \quad & \lesssim \,  \frac{\pi^2 g_\ast}{30} \, \frac{T_\mathrm{EW}^4}{v_\Phi^2} \, \nonumber \\ & \sim \, \left( 10 \,\text{GeV}\right)^2  \left(\frac{g_\ast}{100}\right) \left( \frac{10^4 \,\text{GeV}}{v_\Phi}\right)^2 \, .
\end{align}
Then, requiring $v_\Phi \gtrsim 10^4$ GeV, as motivated in \Cref{sec:flavObs}, we see that $m_0 \lesssim 10$ GeV. However, a stronger constraint comes from the supercooling. At leading order around $\Phi = 0$, the tree level potential in combination with the finite temperature corrections can be expressed as
\begin{equation}
    V_0 + V_T \sim \left(y_\xi^2 T^2 - \mu^2\right) \Phi^2
\end{equation}
and since the flavon mass is given by $m_0^2 = 2 \mu^2$, large flavon masses imply earlier FNSB. Additionally, the CW contribution when $y_\xi$ is too large ($ \sim 1$) prevents a minimum from forming in the correct range, and, carrying out a numerical scan tracking all contributions, we find feasible points with a vev in the correct range ($\sim 0.1 \Lambda_\mathrm{FN}$) and SB delayed until after EWSB only for 
\begin{equation}
    m_0^2 \lesssim 1 \, \mathrm{GeV} \, ,
\end{equation}
incompatible with the constraints from lepton flavor violation. Furthermore, the found minima are only metastable. After EWSB, the CW contribution of the SM fermions drives the potential to negative values at large $\Phi$. Neither restriction applies when including higher dimensional operators in the potential.
\section{Flavor bounds}\label{app:flavorbounds}
To determine the contribution to flavor changing observables from the light flavon, the mass matrix
\begin{equation}
    m^f_{ij} = \frac{v}{\sqrt{2}} y_{ij}^f \epsilon^{n^f_{ij}}
\end{equation}
has to be diagonalised as
\begin{equation}
    \left( U_L^f\right)^\dagger m^f U_R^f = \hat{m}^f \, .
\end{equation}
Although this diagonalises the Higgs couplings, the flavon interactions retain off-diagonal entries
\begin{equation}
    \mathcal{L} \supset c_{ij} \bar{f}_{L_i} (\rho + i a) f_{R_j}  + h.c.
    \label{eq:FCinteractions}
\end{equation}
with coupling matrix
\begin{equation}
    c = \frac{1}{v_\Phi}  \left( U_L^f\right)^\dagger \left( m^f n^f\right) U_R^f \, ,
\end{equation}
and these induce the flavor changing processes characteristic of FN frameworks. The couplings of radial and axial mode differ by a factor of $i$.
\subsection{ALP contribution to $\mu \rightarrow e \gamma$}
\begin{figure}
    \centering
    \includegraphics[width=\linewidth]{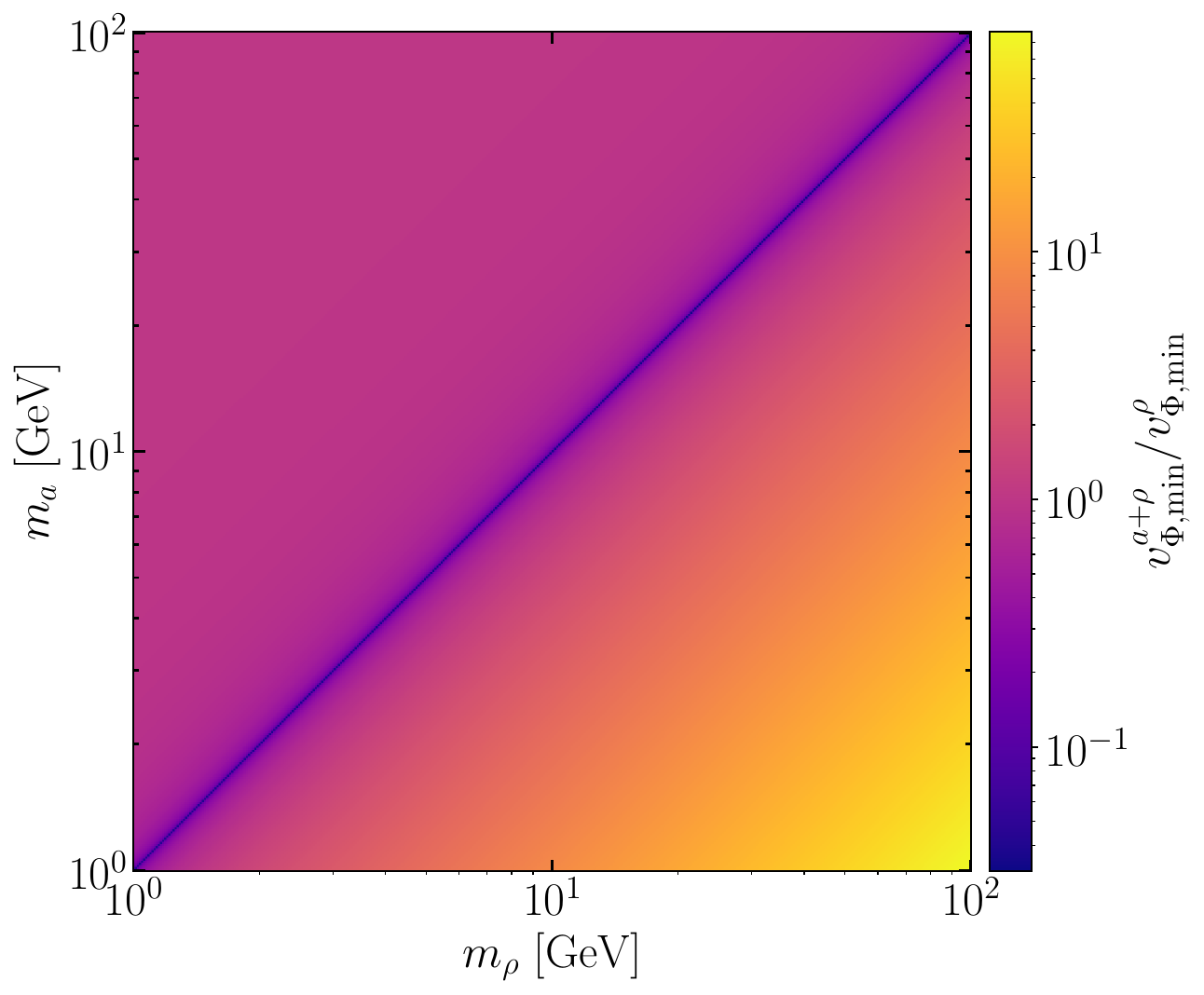}
    \caption{The color-coding shows the $\mu \rightarrow e \gamma$ bound on the flavon vev as a ratio between the estimate in \Cref{sec:flavObs} when only the radial mode is taken into account, compared to the value when both radial and axial mode are explicitly included. See text for details.}
    \label{fig:ALPcontrFlavor}
\end{figure}
In the main text, we took the ALP mass as a free parameter and therefore did not explicitly include its contribution to the flavor observables, as it can both increase and decrease the constraints. \Cref{fig:ALPcontrFlavor} shows the ratio 
\begin{equation}
    r = \frac{(v_\Phi^\textrm{min})_{\rho+a}}{(v_\Phi^\textrm{min})_{\rho}}
\end{equation}
between the minimal allowed vev $v_\Phi^\textrm{min}$ when both fields are taken into account and only for the radial mode. Changing the couplings in \cref{eq:Cij_e} by (i), the term in the second line obtains a minus sign and therefore, for similar radial and axial masses, the contributions to $\mu \rightarrow e \gamma$ show some cancellations. If one field is lighter than the other, it gives the leading constraints on the flavon vev, which can be understood from the plot as follows: when $m_a > m_\rho$, the ratio remains at 1 while for lighter ALPs the ratio grows larger.  
\subsection{Bounds from the quark sector}
\begin{figure}
    \centering
    \includegraphics[width=\linewidth]{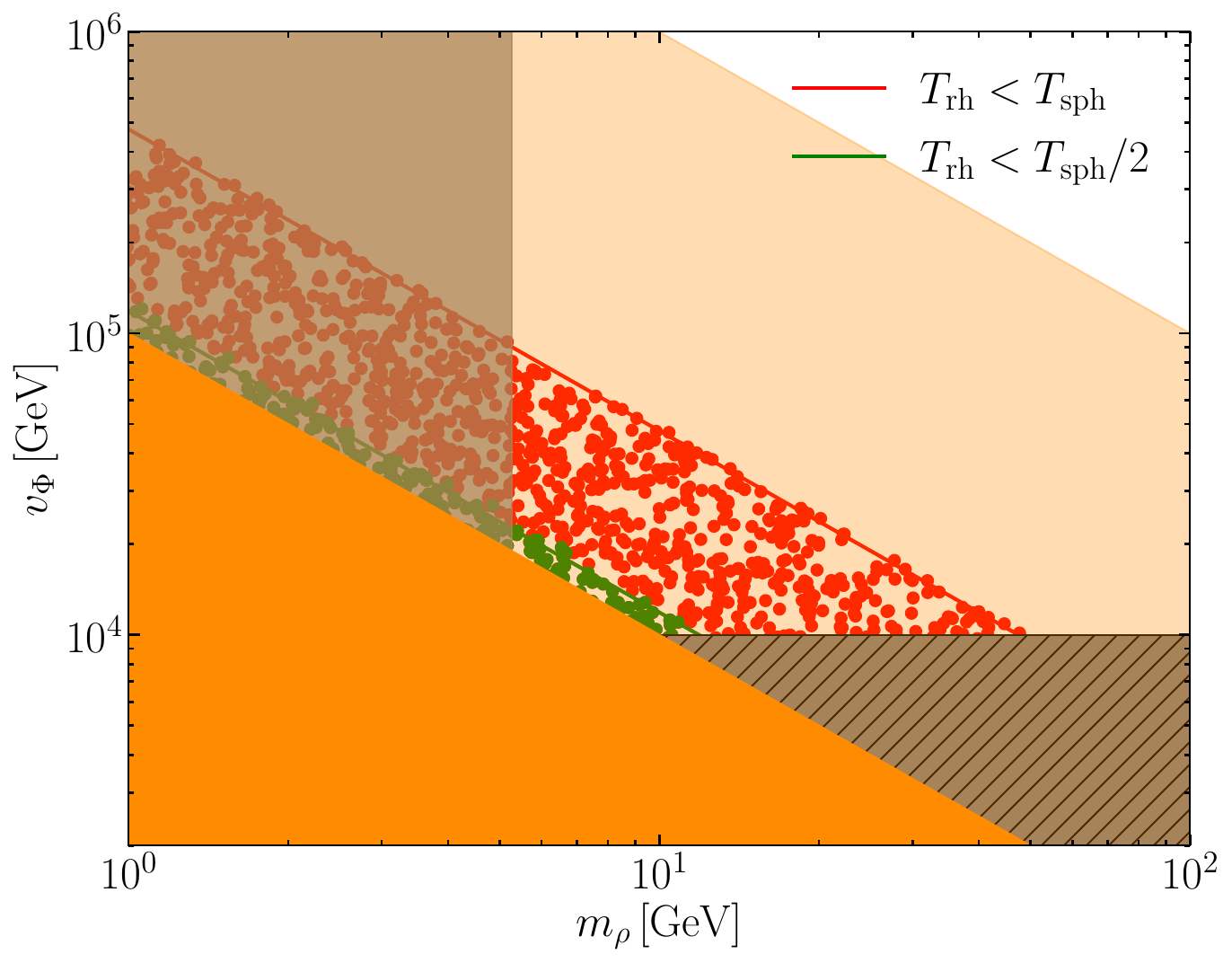}
    \caption{FCNC bounds from $K-\overline{K}$ oscillations, naively (light orange) 
    or with texture suppression (orange). Low masses are not included due to very stringent invisible meson decay bounds. The excluded lower region comes from $\mu \rightarrow e \gamma$, see \Cref{sec:flavObs}. The colorful lines and points showcase maximal reheating temperatures for the viable points, see for details Sec.\ref{sec:PT}.}
    \label{fig:constraintsMassVev_KaonViaTexture}
\end{figure}
We separate the constraints that come from flavor changing processes in the quark sector into two regimes, depending on whether the flavon is heavier or lighter than the kaon and B-meson mass. Only the processes leading to the dominant constraints are listed, see for instance \cite{Greljo:2024evt} for a more complete overview over all flavor observables.
\subsubsection{Invisible meson decays}
When $\rho$ is very light, invisible meson decays become kinematically allowed which are tightly constrained, see e.g. Ref \cite{Greljo:2024evt}. Specifically, when $m_\rho < m_K \,(m_B) \sim 0.5\,\rm{GeV} \,(5\,\rm{GeV})$, the invisible decays $K \rightarrow \pi \rho$ and $B \rightarrow  K^{(\ast)} \rho$ require the flavon vev to be larger than compatible with the reheating constraints determined in \Cref{sec:PT}.\footnote{There is only a narrow $m_\rho - v_\Phi$ band allowed for which the ALP decays inside the detector, see for details \cite{Greljo:2024evt}.} Therefore, when showing the constraints on the flavon vev in \Cref{fig:constraintsMassVev_KaonViaTexture} we have narrowed the mass range compared to \Cref{fig:Leptonic sector}.
\subsubsection{Meson mixing}
For heavier scalars, the dominant constraints come from meson mixing $P^0 - \overline{P}^0$, particularly from kaon mixing. As for $\mu \rightarrow e \gamma$, the precise constraint depends on the FN charge assignments which induce the off-diagonal quark couplings. Indeed, doing a naive analysis similar to the one in Ref.\,\cite{Baldes:2016gaf}, and comparing the UTfit constraint \cite{UTfit:2026qxi}
\begin{equation}
    \left| \rm{Im}[C_4^{sd}]\right| \lesssim 4 \times 10^{-18} \, \rm{GeV}^{-2}
\end{equation}
with the expected Wilson coefficient
\begin{equation}
    C_4^{sd} \sim \left(\epsilon^{n_{sd}-1} \frac{v_h}{2 \Lambda_{\rm FN}} \frac{1}{m_\rho} \right)^2 \, ,
\end{equation}
where $n_{sd}$ is the sum of charges of the down-strange mixing terms, which might e.g. be $n_{sd} \sim 4$ with $\epsilon \sim 0.1$, we find
\begin{equation}
    v_\Phi \gtrsim 10 \, \rm{TeV} \times \left( \frac{1 \, \rm{TeV}}{m_\rho}\right) \quad \textbf{[naive]}.
\end{equation}
However, as shown in \cite{Cornella:2023zme}, the naive bound can be up to two orders of magnitude higher than strictly required. A favorable charge assignment can thus improve the bound by a factor 100 to be roughly, matching what was also found in \cite{Greljo:2024evt},
\begin{equation}
    v_\Phi \gtrsim 10 \,\rm{TeV} \times \left( \frac{10 \, \rm{GeV}}{m_\rho}\right) \quad \textbf{[best case]} \, .
\end{equation}
\Cref{fig:constraintsMassVev_KaonViaTexture} shows the constraints in the quark sector, with the naive bound in light orange and the stringent bound in orange. The mass range is cut off at 5 GeV, due to the invisible meson decays (transparent gray cut-off in the plot). This, in combination with $T_{\rm rh}<T_{\rm sph}$ (red line), leads to an upper bound on the VEV of $\sim 10^5$ GeV. With $\epsilon \sim 0.1$, this can be translated to an upper bound on $\Lambda_\textrm{FN}\sim 10^6$ GeV. Comparing with \Cref{fig: T_LR Lambda_F}, it becomes clear that with such a low UV scale for a lepton with FN charge $1$ (as is expected for the tau) there is no temperature range in which equilibration has become ineffective while sphalerons are active.

\bibliography{bib.bib} 

\end{document}